%% file: main.tex
\documentclass[letterpaper,twocolumn,10pt]{article}
\usepackage{usenix-2020-09}

\usepackage{tikz}
\usepackage{amsmath}
\newcommand{\NetApps}{MNCs\xspace}
\newcommand{\netApp}{MNC\xspace}
\newcommand{\netApps}{MNCs\xspace}

\usepackage{filecontents}
\input{sections/0-macro}

\usepackage{diagbox}
\usepackage{empheq}
\usepackage[subtle]{savetrees}
\begin{document}

\date{}

\title{\Large \bf Robustifying ML-powered Network Classifiers with PANTS  
}

\author{
{Minhao Jin}\\
Princeton University
\and
{Maria Apostolaki}\\
Princeton University
} 

\maketitle
\pagestyle{plain}

\begin{abstract}
    \input{sections/0-abstract}
\end{abstract}

\input{sections/1-intro}

\input{sections/2-requirement}
\input{sections/3-overview}

\input{sections/4-design}

\input{sections/5-robustification}
\input{sections/6-eval}
\input{sections/8-conclusion}

\input{sections/acknowledgement}
\input{sections/ethics}
\input{sections/open-science}
\bibliographystyle{plain}
\bibliography{reference}

\input{sections/appendix}

\end{document}

%% file: sections/0-macro.tex
\usepackage{multirow}
\usepackage{mathtools}
\usepackage{xspace}
\usepackage{xcolor}
\usepackage{multirow}
\usepackage{graphicx}
\usepackage{booktabs}
\usepackage{tabularx}
\usepackage{array}
\usepackage{caption}
\usepackage{subcaption}
\usepackage{amsmath}
\usepackage{bm}
\usepackage{pifont}
\newcommand{\cmark}{\ding{51}}%
\newcommand{\xmark}{\ding{55}}%
\usepackage{url}
\usepackage{xurl}
\usepackage{makecell}

\usepackage{algorithm,algpseudocode}
\algnewcommand{\To}{\textbf{To }}
\algnewcommand\Input{\item[\textbf{Input:}]}%
\algnewcommand\Output{\item[\textbf{Output:}]}%
\algrenewcommand{\algorithmiccomment}[1]{%
  \hfill\(\triangleright\)\textit{\tiny #1}%
}

\usepackage{mathtools}

\newcommand{\todo}[1]
{{\footnotesize\color{red}[ToDo: #1]}}

\newcommand{\new}[1]
{{\color{black}#1}}
\newcommand{\change}[1]
{{\color{black}#1}}
\newcommand{\final}[1]
{{\color{black}#1}}

\newcounter{packednmbr}

\newcommand{\mypara}[1]{\smallskip \noindent{\bf {#1}.}~}

\newcommand{\remove}[1]{}

\newcommand{\ie}{\emph{i.e.,}\xspace}
\newcommand{\eg}{\emph{e.g.,}\xspace}
\newcommand{\etc}{\emph{etc.}\xspace}

\newcommand{\sys}{PANTS\xspace}

 \newcommand{\myitem}[1]{\vspace*{0.02in}\noindent\textbf{#1}}

\usepackage{ifthen}

\usepackage{listings}
\newcounter{insightlabel}
\newcounter{insightnmbr}
\renewcommand{\theinsightlabel}{\textbf{\theinsightnmbr}}

\newcounter{challengelabel}
\newcounter{challengenmbr}
\renewcommand{\thechallengelabel}{\textbf{\thechallengenmbr}}

\newcounter{researchquestionlabel}
\newcounter{researchquestionnmbr}
\renewcommand{\theresearchquestionlabel}{\textbf{\theresearchquestionnmbr}}

\newcounter{findinglabel}
\newcounter{findingnmbr}
\renewcommand{\thefindinglabel}{\textbf{\thefindingnmbr}}
\newenvironment{finding}{
\begin{list}{\textbf{Finding }\thefindinglabel:~}{\usecounter{findinglabel}\stepcounter{findingnmbr}\setlength{\labelwidth}{0pt}\setlength{\labelsep}{0pt}\setlength{\leftmargin}{0in}\noindent\rule{\linewidth}{1pt}\vspace{-8pt}\item \bf \em}}{\\[-7pt]\end{list}\vspace{-10pt}\noindent\rule{\linewidth}{1pt}}

\newcounter{limitationlabel}
\newcounter{limitationnmbr}
\renewcommand{\thelimitationlabel}{\textbf{\thelimitationnmbr}}

\newcommand{\commentout}[1]{}

\newcommand{\rom}[1]{\uppercase\expandafter{\romannumeral #1\relax}}

%% file: sections/0-abstract.tex
Multiple network management tasks, from resource allocation to intrusion detection, rely on some form of ML-based network traffic classification (\netApp). 
Despite their potential, \netApps are vulnerable to adversarial inputs, which can lead to outages, poor decision-making, and security violations, among other issues. 

The goal of this paper is to help network operators assess and enhance the robustness of their \netApp against adversarial inputs.
The most critical step for this is generating inputs that can fool the \netApp while being realizable under various threat models. 
Compared to other ML models, finding adversarial inputs against \netApps is more challenging due to the existence of non-differentiable components \eg traffic engineering and the need to constrain inputs to preserve semantics and ensure reliability. These factors prevent the direct use of well-established gradient-based methods developed in adversarial ML (AML).

To address these challenges, we introduce \sys, a practical white-box framework that uniquely integrates AML techniques with Satisfiability Modulo Theories (SMT) solvers to generate adversarial inputs for \netApps. We also embed \sys into an iterative adversarial training process that enhances the robustness of \netApps against adversarial inputs.
\sys is 70\% and 2x more likely in median to find adversarial inputs against target \netApps compared to state-of-the-art baselines, namely Amoeba and BAP.
\new{\sys improves the robustness of the target \netApps by 52.7\%  (even against attackers outside of what is considered during robustification)
without sacrificing their accuracy. }



%% file: sections/1-intro.tex
\vspace{-0.4cm}
\section{Introduction}
\label{sec:intro}


Managing networks involves tasks such as performance routing, load balancing, and intrusion detection and is extremely challenging due to the highly diverse, heavy-tailed, and unpredictable inputs, \ie network traffic. 
ML-based network-traffic classification (\netApps) provides a compelling alternative for handling these complex tasks more efficiently~\cite{utmobile,sharma2023vca,iscxvpn2016,sharma2022lumen,kumar2022iot,Doshi2018machine,jiang2016cfa,jiang2016via,yao2021lb}. However, like other ML-based solutions~\cite{5504793, goodfellow2014explaining}, \netApps are vulnerable to adversarial inputs, that is, inputs meticulously perturbed to deceive the \netApp.  Attackers can exploit this to compromise critical network infrastructure by subtly perturbing traffic characteristics like packet sizes and inter-arrival times, leading to security breaches or downtime. Beyond being dangerous, these attacks are particularly practical in the networking domain, where multiple entities have access to network traffic.

What is critically lacking is a systematic framework for network operators and \netApp designers to enhance the robustness of \netApps against attackers. 
This involves identifying adversarial inputs, meaning perturbations on original packet sequences that deceive the target \netApp into misclassifying them. 
Operators can then use these inputs to:
\emph{(i)} refine and optimize the \netApp by exploring different models, architectures, or features; \emph{(ii)} assess the attack surface of their \netApps, understand the potential consequences and plan accordingly; and \emph{(iii)} use them to strengthen \netApps against realistic threat models through re-training or fine-tuning.
Critically, operators will not use or take the framework seriously if the generated adversarial inputs are  \emph{(i)} not realizable, meaning they do not fall within a realistic attacker's capabilities, or  \emph{(ii)}  do not retain the semantics of the original sequence, meaning the ground-truth label of the perturbed inputs has changed.
\change{More importantly, we show that adversarial inputs that are non-realizable or non-semantics-preserving can hurt the model's accuracy when used in training.} 

\remove{
\begin{figure}[hbt!]
\centering
    \includegraphics[width=.95\linewidth]{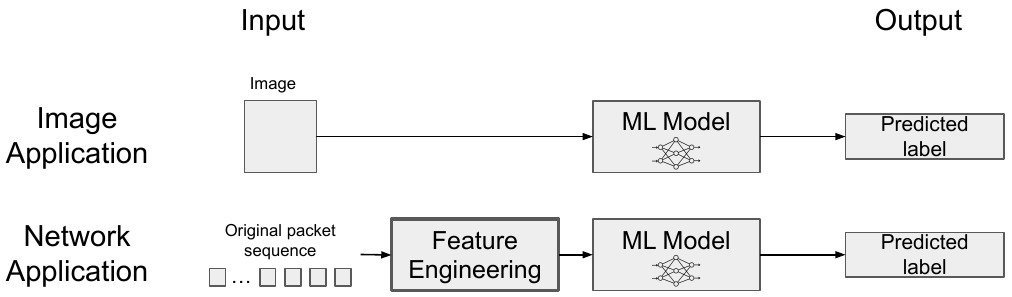}  
    \caption{Image applications usually apply an end-to-end ML model for classification while networking applications commonly include an additional non-differential and non-invertible feature engineering module. }
    \label{fig:diff}
\end{figure}
}

While there are numerous tools for generating adversarial samples in the Adversarial Machine Learning (AML) literature~\cite{goodfellow2014fgsm,madry2017pgd,chen2017zoo,carlini2017towards,wang2023adversarial} they are not directly applicable to \netApps. 
First, AML tools are built for systems that are end-to-end differentiable, which makes them unsuitable for typical \netApps that include a feature engineering module to transform packet sequences into feature vectors before they are fed to the ML model. 
Although AML tools can generate adversarial feature vectors, there is no guarantee these vectors can be reverse-engineered into valid packet sequences. Even if a valid packet sequence exists, the perturbations needed to convert the original sequence into an adversarial one may exceed what a realistic attacker could achieve.
Second, it is uncertain whether AML-generated samples are preserving the semantics of the original packet sequence while deceiving the \netApp. Although AML techniques often incorporate a perturbation budget (expressed as a norm), this concept does not translate well to networking traffic. For example, perturbing a VoIP flow (packet sequence) to resemble web browsing by delaying packets might deceive a traffic classifier but degrades the original flow's performance, making the call unusable.
Critically, adversarially training an \netApp with non-realizable or non-semantics-preserving samples (\ie directly using what AML generates) will degrade the \netApp's accuracy without significantly improving robustness as we show in \S\ref{subsec:findings}.

Learning-based approaches such as Amoeba~\cite{amoeba} use reinforcement learning to directly generate adversarial traffic sequences. Being black-box approaches, such tools are great for attackers to fool ML-powered applications. However, the lack of diversity in adversarial samples, their instability, and training hardness make them less effective for assessing and enhancing the robustness of \netApps from the network operator's perspective, as we show in our evaluation.

To address these shortcomings, this paper presents \sys, a
framework to generate Practical Adversarial Network Traffic Samples that are realizable for various threat models and semantic-preserving. \change{\sys also enhances the target \netApp through a novel iterative augmentation training process that leverages these samples.}
\sys integrates traditional methods for generating adversarial samples (AML), such as Projected Gradient Descent (PGD)~\cite{madry2017pgd} or Zeroth-order Optimization (ZOO)~\cite{chen2017zoo},  with formal methods, concretely a Satisfiability Modulo Theories (SMT) solver. The AML perturbs the original inputs (packet sequences or features) in a manner that maximizes their distance  (\eg loss for MLP) from the original decision boundaries with a consistently high success rate. 
The SMT solver finds packet sequences that do not diverge too much from the AML's output but are realizable, consistent with the threat model and preserve the semantics of the original packet sequence. 
To achieve this, we encode the feature engineering, the threat model, and other networking constraints needed for semantic preservation into logical formulas. Among other optimizations, \sys iteratively adds and removes constraints that come from the AML, prioritizing those that have a substantial effect on confusing the classifier.
Finally, we integrate this generative process into an interactive training process that enhances the target \netApp.

We evaluated \sys against three distinct ML-powered network classifiers, each implemented with different models and processing pipelines. \sys finds adversarial, realizable and semantic-preserving samples with 35.31\% success rate in median, that is 70\% and 2x higher than two SOTA methods. 
Despite being a white-box approach, \sys works well with gradient (\eg transformer encoder) and non-gradient (\eg random forest) models and various processing pipelines, including those containing non-differentiable components (\eg feature engineering). 
\change{Importantly, \sys enhances \netApps's robustness against both white-box and black-box attackers, even when their capabilities differ from or exceed those considered during the robustification process.} 

We will open source \sys and our evaluation materials to facilitate further research and \netApps robustness
(cf.~\ref{sec:open-science}).

%% file: sections/2-requirement.tex
\vspace{-0.25cm}
\section{Motivation \& Limitations of Existing Work}\label{sec:motivation}
\begin{figure}[h]
\centering

\begin{subfigure}{.60\linewidth}
    \centering
    \includegraphics[width=.95\linewidth]{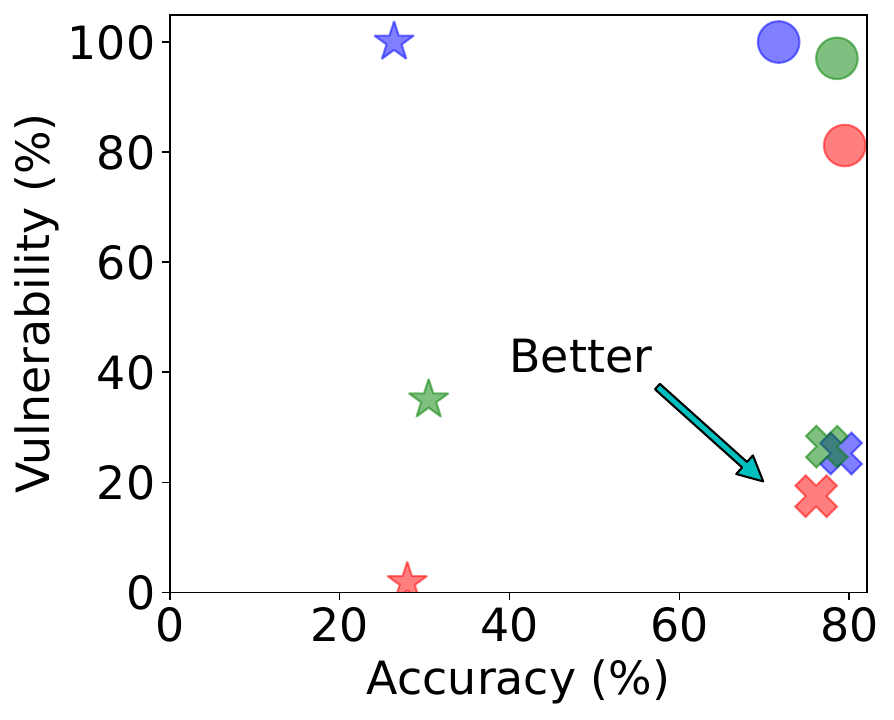} 
    \label{motivation-pants_1}
\end{subfigure}
\begin{subfigure}{.35\linewidth}
    \centering
    \includegraphics[width=.99\linewidth, ]{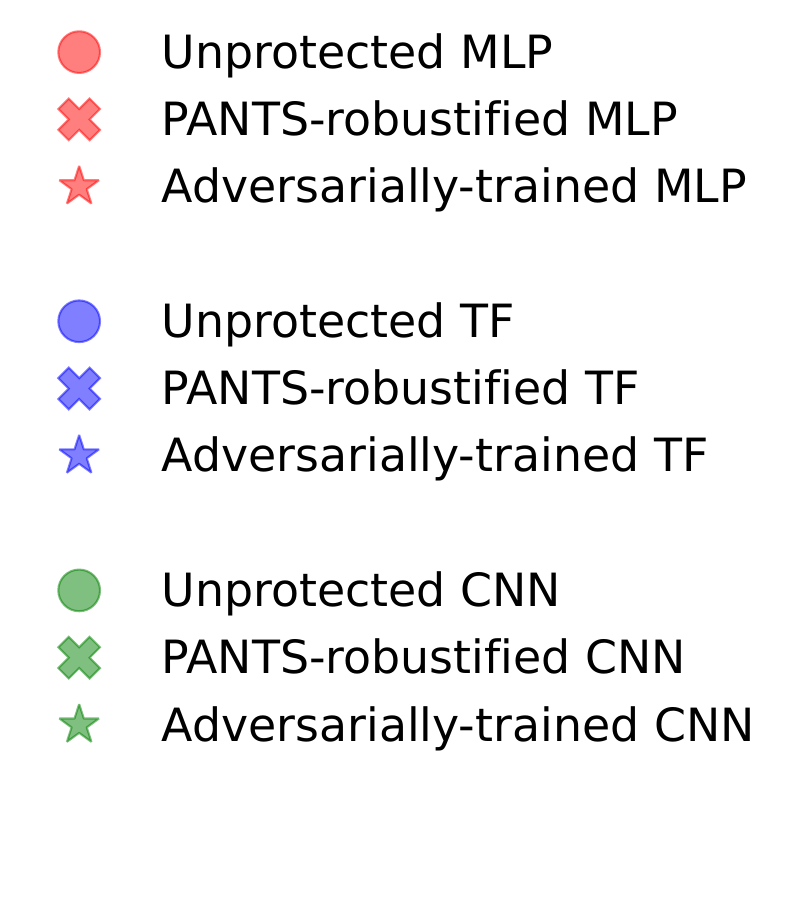}  
\end{subfigure}
 \caption{\netApps are vulnerable to adversarial inputs (circles). Traditional adversarial training sacrifices accuracy to reduce vulnerability (star). \sys iterative adversarial-training reduces the vulnerability without hurting accuracy (cross). }
    \label{motivation-pants}
\end{figure}

To highlight the need for a framework that generates adversarial inputs to help operators assess and mitigate the vulnerabilities of their \netApps, we present a specific use case: an ML-based classifier that exposes the network to a range of exploits, from giving attackers an unfair advantage to causing outages, regardless of the exact implementation of the \netApp. We then outline the essential properties required for such a framework and explain why existing solutions are insufficient.

\subsection{Motivating use case}

Consider a network operator that uses an ML-based traffic classifier to distinguish real-time applications (\eg online gaming). The operator uses this information to optimize routing: real-time applications use fast but low-bandwidth network paths.
This is one of the multiple network management tasks for which network operators could use an ML-based traffic classifier (built with Multilayer Perceptron-MLP, Random Forest, Transformers~\cite{lin2022bert,Aneja2023iot, utmobile,sharma2023vca,iscxvpn2016}) and benefit from its ability to learn complex patterns directly from data. 
While highly useful, this \netApp exposes the network to various exploits. 
Two representative examples include attackers controlling \emph{(i)} the end hosts of the communication or \emph{(ii)} an in-network component.
In the former case,  a malicious (or simply ``rational'') application developer could manipulate the traffic generated by her application to be misclassified as gaming by the network's \netApp, hence gaining an unfair performance advantage over competitors. This could be achieved by subtly altering the size of some packets or delaying specific packets without significantly impacting the overall performance of her application's traffic.
In the latter case, a malicious or compromised router in-path can increase the volume of traffic routed through the fast but low-bandwidth path by subtly altering the non-real-time traffic to cause the \netApp to misclassify it as real-time. These perturbations would result in congestion in the low-bandwidth path, leading to delays or even downtime for actual real-time traffic, which is vital for the network's profitability. 

\myitem{ML network classifiers are vulnerable to adversarial inputs.}
Fig.~\ref{motivation-pants} illustrates the likelihood of an attacker controlling end hosts (e.g., a malicious app developer) to successfully manipulate their traffic to cause an \netApp to misclassify it (predict a favorable for the attacker label). 
The results show that, despite the high accuracy of the models during testing, the attacker can deceive classifiers implemented with MLP, CNN, and a transformer with success rates of 81.12\%, 94.25\% and 99.80\%, respectively.
While we identified these manipulations (adversarial inputs), hence the vulnerability of the \netApps using \sys, which is a white-box approach with complete access to the \netApp's implementation, this does not diminish the significance of the risk they pose.
First, security through obscurity is never a sound strategy, and network operators should be prepared for (or at least aware of) the worst-case scenario where an attacker has complete knowledge of their system.
\change{Second, black-box approaches~\cite{amoeba} are also able to find examples that fool \netApps, albeit with less likelihood.
That is to show that adversarial examples are not random bugs but stem from learning from non-robust features, \ie relying on unrelated co-occurrences in data instead of true correlations~\cite{ilyas2019adversarial}. 
}

\myitem{Adversarial training trades accuracy for vulnerability.}
Fig.~\ref{motivation-pants} also shows the trade-off between inaccuracy and vulnerability in \netApps. Indeed, an operator could leverage authentic (off-the-shelf) adversarial training to shield an \netApp against adversarial inputs. However, by doing so, she will sacrifice the accuracy of the \netApp, meaning the baseline performance of the \netApp in more common inputs will degrade. 
By authentic adversarial training, we refer to training a robustified model with the help of PGD ~\cite{madry2017pgd}. During model training, the original training samples are substituted with their PGD-generated adversarial counterparts in each training iteration.

\subsection{Requirements}
\label{sec:requirement}
Next, we elaborate on the requirements of a framework that generates adversarial samples and use them to enhance the robustness of \netApps before we explain why existing approaches fall short.


\mypara{Adversariness \& Semantic preservation }
The framework needs to generate inputs that are adversarial, meaning that they are capable of misleading \netApps into producing an incorrect classification.
To ensure that the classification is indeed incorrect, these inputs should be created by meticulously applying perturbations to the original labeled samples without altering their semantics. In other words, each adversarial input should still represent the same underlying meaning or class as the original, which is described as a set of constraints on the adversarial input.
The closest notion of semantics in the image domain is the perturbation budget, which sets a maximum allowable distance between the original and perturbed image~\cite{madry2017pgd}. However, such a metric doesn't translate directly to the networking domain, where there are dependencies across packets or features of each input, which perturbations should respect. We assume that semantics can be expressed as constraints on the variables of the input (not only as a distance between the adversarial and the original), and the framework should provide the means to respect them.


\mypara{Realizability} The framework should provide mechanisms to constrain the generated adversarial inputs according to domain-specific rules and a configurable threat model, which the operator will consider reasonable. Concretely, adversarial samples must represent valid packet sequences — for example, TCP should never acknowledge a packet before it is sent.
Further, each adversarial input must be created by applying perturbations that are within the capabilities of the threat model to an original input. Optimizing the model by training it on unrealizable inputs that it will never encounter is not only wasteful but can also degrade its performance on common inputs, as we show in \S\ref{subsec:findings}. For instance, an in-network attacker cannot alter an encrypted packet in a way that evades detection, so examining such scenarios would be unproductive.

\mypara{Generality} 
The framework needs to be able to work in various classification pipelines.  Unlike the image domain, where end-to-end neural network-based (NN-based) models play a dominant role in various tasks, networking applications employ a wide spectrum of classification models and pipelines. Concretely, the framework needs to work for typical \netApps, which include non-differentiable and non-invertible feature engineering modules to extract statistical features from the given packet sequence, followed by a differentiable or non-differentiable ML model such as Multi-layer perception (MLP) or Random Forest (RF) for classification. 
The framework should also work for \netApps that directly encode the packet sequence into a sequence of packet length, packet direction, and inter-arrival times only. In the latter case, the encoded sequence is fed into deep learning models such as Transformer (TF) or convolutional neural network (CNN) for classification. 


\remove{
\begin{table}[hbt!]
\centering
\scalebox{0.95}{
\begin{tabular}{c|ccc}
 & \small{Synthetic} & \small{RL} & \small{Gradient} \\ \hline
\parbox[c]{3cm}{\centering \small{Adversariness \& \\ semantics preservation}} & \xmark & \small{Sometimes} & \xmark \\
\small{Realizability} & \xmark & \cmark & \cmark \\
\small{Generality} & \small{N/A} & \cmark & \xmark
\end{tabular}
}
\caption{Existing approaches for generating adversarial samples that could be used for robustifying ML-powered networking apps fail to meet our requirements.\todo{update}}
\label{tab:requirement}
\end{table}
}
\subsection{Limitations of existing work}
\label{subsec:limit_existing_work}
Having described the requirements of a framework that aims to help network operators, we explain why various lines of work that generate synthetic or adversarial \emph{networking} inputs fail to satisfy them. 
We elaborate on works that generate adversarial images (\ie AML) in \S\ref{subsec:aml-promise}.

\myitem{Synthetic data generation}
%
\new{ Networking-specific generative models, such as NetShare~\cite{netshare-sigcomm2022} and NetDiffusion~\cite{netdifJiang}, have demonstrated strong potential in generating synthetic traffic. However,  synthetic traffic is not adversarial—\ie it should not mislead the downstream applications such as an \netApp.
In fact, such works aim to generate samples that match the original distribution (per the fidelity definition), whereas adversarial samples’ distribution is typically different from the training distribution~\cite{carlini2017towards}.
 This distinction is critical, as training with adversarial samples has been shown to produce more robust models compared to merely increasing the size of the training dataset~\cite{madry2017pgd}.
 As a result, NetShare~\cite{netshare-sigcomm2022} is 92\%(99\%, 92\%) less effective in robustifying RF (TF, CNN) compared to \sys and cannot robustify MLP, unlike \sys, as we show in Fig.~\ref{fig:netshare_robustification} and Fig.~\ref{fig:apdx:netshare_robustification}.
These works may also face a semantics preservation issue: synthetically generated traces from original malicious traces (\eg shrew attacks) might not achieve the malicious intent (\eg cause an outage); hence might not be truly malicious.

}

\myitem{Black-box RL-based techniques} 
While useful for aspiring attackers, black-box RL-based approaches~\cite{tomer2019robustify} for generating adversarial examples against \netApps are incapable of meeting our requirements. 
Amoeba~\cite{amoeba}, for example, trains an agent to perturb each packet to maximize the likelihood that the altered flow will deceive the \netApp. Amoeba fulfills the generality requirement as it is black-box, and can fulfill realizability if trained with the right actions. Still, Amoeba's outputs are not always semantic-preserving and adversarial because there is no way to guarantee those during inference. Also, as we show in our evaluation, Amoeba's generated adversarial inputs are less effective in robustifying \netApps during adversarial training, possibly because it (over)exploits a vulnerability rather than trying to find new ones. \change{Ultimately, this strategy benefits attackers but not operators seeking to enhance the \netApp.}

\myitem{White-box gradient-based techniques} \change{Excluding AML, which is discussed in detail in \S\ref{subsec:aml_limit}, BAP~\cite{BAP} stands out as a notable white-box approach for adversarial generation, particularly due to its focus on networking inputs.}
 BAP  adversarially trains a generator to create perturbation on every single packet of a packet sequence \change{(\eg delay, add dummy bytes)} such that the distance (\eg loss) between the predicted label and the ground truth is maximized. Because BAP uses gradient for maximizing the distance, it does not work for non-end-to-end differentiable \netApps, hence failing the generality requirement. 
Further, by perturbing each packet independently, BAP cannot enforce semantics, which can be flow-wide. We compare against BAP in \S\ref{sec:eval}.

\myitem{Robustness certification} 
Ideally, network operators would resort to robustness certification~\cite{li2023sokcertified,wang2021betacertified,xu2020automaticcertified,tjeng2017evaluatingcertified,yang2020randomizedcertified} to understand the vulnerability of their \netApps. Applying such techniques in networking is not trivial, 
BARS~\cite{wang2023bars}, for instance, considers the robustness guarantee for perturbations at the feature level instead of the guarantee at the packet level and hence cannot incorporate threat models. Indeed, a small perturbation in the flow, such as injecting a dummy packet in a small interarrival time, can easily exceed its robustness region in the feature space. 
More importantly, though, certified robustness cannot match the empirical robustness using adversarial training~\cite{wong2020fast, nandy2021towards}, which is, in fact, more indicative of the real risk.

%% file: sections/3-overview.tex
\vspace{-0.25cm}
\section{Overview}
\label{sec:overview}
Our goal is to build a framework that generates adversarial samples that can be used to assess and improve the robustness of an \netApp while fulfilling the requirements in \S\ref{sec:requirement}.
After we formulate our problem, we elaborate on the opportunities and challenges of traditional AML in solving it. Finally, we explain the insights that drove our design.

\begin{figure}[h]
\centering
     \includegraphics[width=.95\linewidth]{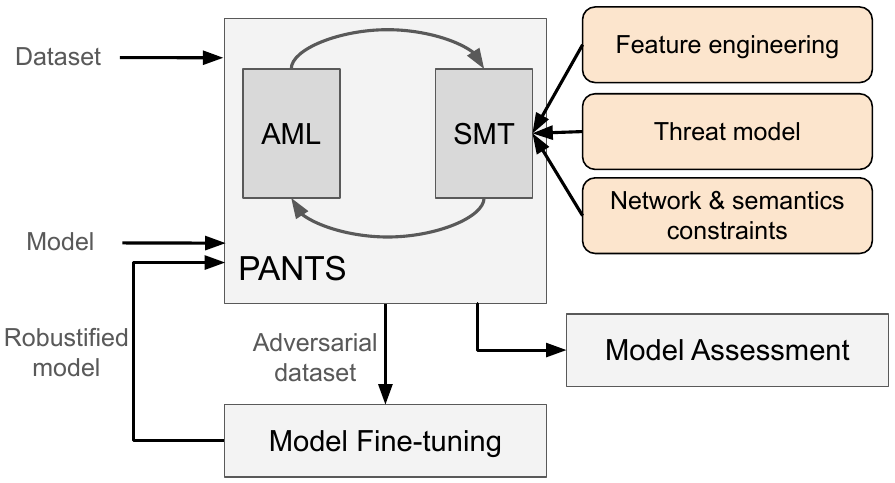}  
\caption{Overview of \sys workflow. \sys generates adversarial inputs that are also used to iteratively train the target \netApp. \sys receives the implementation of a \netApp together with a training dataset and a couple of rules that constrain the generated inputs. At its core, \sys features an AML component that collaborates with an SMT
solver.}
\label{fig:requirement}
\vspace{-0.35cm}
\end{figure}

\subsection{Problem formulation}

Consider an ML model $f$ used to implement a network classifier 
\(f:\mathcal{R}^n \rightarrow \mathcal{Y}\)
where \( \mathcal{Y} \) represents the set of possible labels. 
We define \( {x}  \in \mathcal{R}^n \) as the input to $f$. This can be the result of a feature engineering function, denoted as \( \mathbf{\phi} \) on a sequence of packets p= \( \{p_1, p_2, \dots, p_k\} \) from the set \( \mathcal{P} \) where each \( p_i \) is a packet arrived at time \( i \) or a subset of the sequence of packets p. 
Each packet can be represented by various attributes, including its size flags, direction \etc 
We assume that for every \netApp, the following dataset \( D\) exists.
Let $D = \{ (x_i, y_i) \}_{i=1}^m $ be the dataset, where 
$ x_i \in R^n $ are the feature vectors obtained by applying 
\( \phi \) to raw network data and \( y_i \in Y \) are the corresponding label.\footnote{Applying $f$ directly on a truncated packet sequence is possible and a simplification of the problem we describe.}

We define a threat model as a set of  \emph{perturbations}, $\delta$ 
($\delta: P \rightarrow P $) on the original packet sequences, $p_o$, to generate adversarial packet sequences $p_a$, ($p_a=\delta(p_o) $). 
$\delta(p_o)$ is adversarial if $f(\phi(\delta(p_o)))\neq f(\phi(p_o))$, meaning that the perturbed flow will be misclassified by $f$  and $\delta(p_o)$ preserves the semantics of $p_o$. Semantics are flow-level constraints on $x_o$ \ie constraints that can be dependent on the whole packet sequence, not just each packet independently. 

Our goal is to design a framework that generates packet sequences $p_a$ from $p_o$ such that $p_a$ are adversarial to $f$, semantic-preserving to $p_o$, and are within the capabilities of a threat model.
\new{The framework requires white-box access to $f$, hence is suitable for clouds and ISPs who develop tailored \netApp solutions in-house to eliminate the need to share sensitive information externally (\eg AT\&T~\cite{att}, Azure~\cite{azure}) or third parties developing \netApps (\eg Cisco~\cite{cisco}, Juniper Networks~\cite{juniper}). }

\remove{
Consider an $n$-packets network flow $f$ as a list $f = [p_1^+, p_2^-, p_3^+,..., p_n^+]$ where each packet is $p_i^{\{+, -\}}, i\in[1, n]$. Superscript `$+$' indicates forward packets that are sent from the connection initiator to the target and `$-$' backward packets. 
Each packet $p_i^{\{+, -\}} = (\Delta t_i, l_i)^{\{+, -\}}$ is a tuple where $\Delta t_i$ is the time elapsed between $i$ and $1-1$ packets and $l_i$ is the length of packet. For each flow, packets are ordered by their timestamp in ascending order, \ie $\Delta t_i > 0, i\in[1, n]$. 
Consider, next, an \netApp with a feature engineering module $F$ and an ML model $M$. $F$ extracts statistics (features) from the flow $f$, \eg maximum of inter-arrival times and the mean of packet length. 
After getting $F(f)$, these extracted features are fed into $M$, and $M$ returns its classification result in $o$. The whole process can be written as $o = M(F(f))$. Suppose $o_g$ is sample $f$'s ground truth label, a well-functioning classifier can predict $f$'s label $o$ that $o=o_g$ with high probability. 
}

\begin{figure}[ht]
\centering
\includegraphics[width=0.47\textwidth]{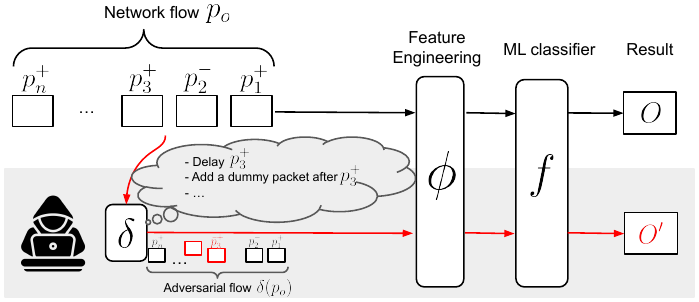}
\caption{Given a network flow p, the perturbation $\delta$ is applied to generate an adversarial flow $\delta(p)$, which causes the ML model to return a wrong output.}
\label{fig:perturb_form}
\vspace{-0.7cm}
\end{figure}

\vspace{-0.1cm}
\subsection{The promise of Adversarial ML}
\label{subsec:aml-promise}
At first glance, adversarial ML (AML) seems to be the right approach for our problem definition and requirements. 
First, \textbf{AML generates adversarial inputs} against ML-based systems, typically in the image domain. 
Concretely, given a trained ML classifier with model parameters $\theta$, an original image sample $x$ with its ground truth label $y$, PGD perturbs $x$ to be an adversarial sample by iteratively running $x^{t+1} = \Pi_{x+S} \left( x^t + \alpha \operatorname{sgn}(\nabla_x L(\theta, x^t, y)) \right)$, $x^0 = x$. $L(\cdot)$ is the target loss function, measuring the distance of the predicted label of $x_t$ and $y$, and $\Pi_{x+S}$ represents the capability of attack, which is usually defined as $\ell_\infty$-ball distance around $x$. The perturbation starts from the original sample $x$. In each iteration, the sample $x^t$ is moved by $\alpha$ following the direction of the gradient of $L$. After modification, the intermediate value is projected to be within the attack capability by clipping it to a predefined $\ell_\infty$-ball distance around $x$. In each iteration, we expect $L(\theta, x^{t+1}, y) > L(\theta, x^t, y)$ to have higher possibility to be misclassified. 
Second, \textbf{AML is quite general}, despite using gradient. In fact, there are AML methods that do not rely on a gradient or whose gradient information is inaccessible to the attacker, such as ZOO~\cite{chen2017zoo}. ZOO uses a combination of techniques, such as coordinate descent and importance sampling, to estimate the gradient for getting adversarial samples. Although it was initially designed for black-box attacks against NN-based models, it can be extended to attack other models such as Random Forest and XGBoost~\cite{Alahmed2022zoorf,artszoorf,artszooxgb}.

\begin{figure*}[h]
\centering
\includegraphics[width=.95\linewidth]{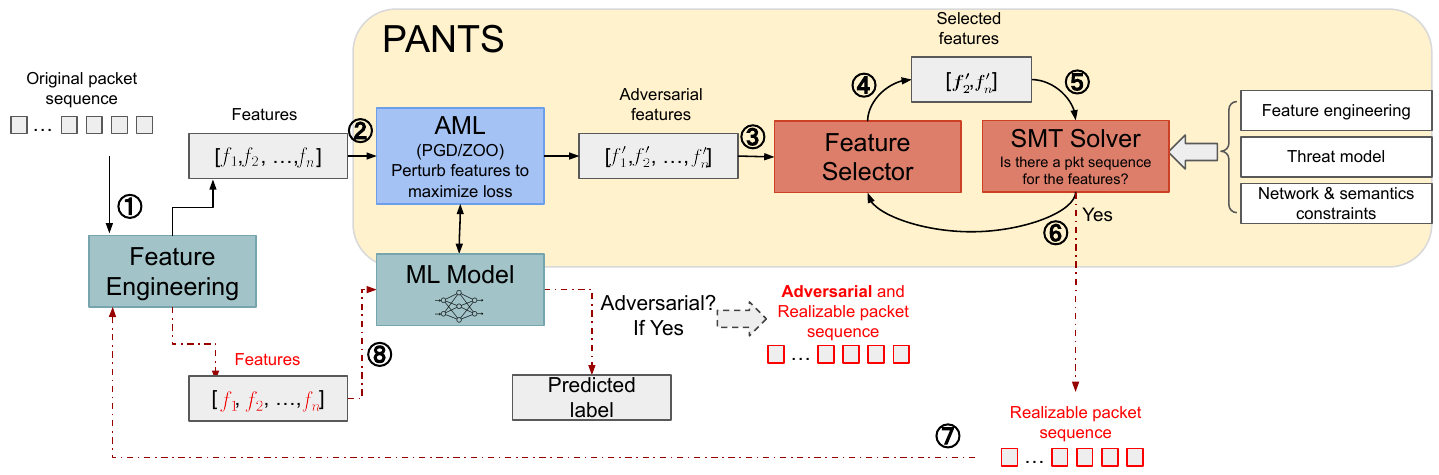}  
\caption{\sys combines AML with an SMT solver to generate adversarial and realizable samples (flows), which are used to assess and enhance robustness.}
\label{fig:e2e}
\end{figure*}

\subsection{Adversarial ML limitations}
\label{subsec:aml_limit}
While AML can generate an adversarial sample and generalize at least across some models, \textbf{AML does not provide realizability and semantics preservation}. 
At a high level, the root cause of the problem is that AML finds adversarial inputs using backward reasoning (\ie targets a change in the predicted label), but semantics preservation and realizability require forward reasoning, \ie are defined on the input~\cite{10.5555/550359,floyd1993assigning}.
In the image domain, this is not a problem as classifiers are end-to-end differentiable: images are directly fed into the ML model for classification, which is typically neural network-based. AML modifies input pixels directly and independently from each other (as they do not need to adhere to any rules beyond being in a range) to increase the loss informed by the gradient. Such techniques also include a budget to constrain just the pixel-to-pixel distance from the original with no other obligation.

However, in the networking domain, the traffic is often processed by the feature engineering module, which generates feature vectors that are fed to the ML model. 
While gradient information can still guide the modification of each feature to increase loss, it is unclear how to map modifications to the feature vector into the corresponding packet sequence as the feature engineering is non-differential and non-invertible. In fact, there is no guarantee that a perturbed feature vector would have a corresponding valid packet sequence as AML changes each feature independently, making contradicting decisions. For example, suppose an ML-powered intrusion detection system whose feature set includes both minimum and maximum packet inter-arrival times ($min\_iats$, $max\_iats$). Since AML does not understand the two features are dependent, in other words, that $min\_iat \leq max\_iats$ must always hold, it is possible for AML to decrease the value of $max\_iats$ and increase $min\_iats$ such that $min\_iats > max\_iats$. In this case, while the generated feature vector might be adversarial in that it might be misclassified by the intrusion detection system, it is not a realizable sample, \ie there is no packet sequence that would ever result in this feature set. In short, using the perturbed feature vectors is like defending against an impossible attack. 
While one can try to use a mask~\cite{peng2019mask} to only allow AML to change a subset of the features that are not dependent, that would reduce the inputs that AML can find. Indeed, while AML can run with as few as a single feature there is no guarantee that the generated sample would be adversarial.


Even if the target model does not use (non-differentiable) feature engineering or if one somehow finds a packet sequence that is consistent with the AML adversarial feature vector, the packet sequence might still fall outside the capabilities of the threat model or violate any network constraints, or break the semantics of the original flow.  
For example, suppose the threat model only allows injecting dummy packets (but not removing), and flow duration $f_1$ and packets per second $f_2$ are two features extracted by the feature engineering module. The number of packets can be derived from $\#packets = f_1/f_2$. If $\#packets$ derived from the adversarial features is less than the number of packets in the original flow, even though one can find an adversarial packet sequence consistent with the adversarial features, it still falls outside the capabilities of the threat model. Similarly, for a deep NN that is directly accepting the packet sequence as input, the AML-generated adversarial sequence could change the interarrivals in a way that is impossible for any congestion control algorithm or in a way that makes the corresponding application unusable. 

\subsection{\sys: Adversarial ML for networking}

To address the inherent challenges in applying AML to \netApps, we integrate a formal-method component to ensure the realizability and semantics-preservation of the produced adversarial samples. 
Our first insight is to use logical formulas to model the feature-engineering process that turns packet sequences into feature vectors and sequence-level constraints for \netApps with no feature-engineering, effectively encoding input dependencies. Further, we encode threat-model capabilities, networking rules and semantics into flow-wide or packet-level constraints and ask an SMT (Satisfiability Modulo Theories) solver to find a packet sequence that satisfies them. If the answer is ``UNSAT'', the adversarial features are not realizable under the given threat model, semantic, and networking constraints. Observe that SMT supports both forward and backward reasoning, hence can connect the requirements for adversariness with realizability and semantics. Critically, this makes \sys extensible to any definition of semantics.

As feature engineering modules and semantics definition can be arbitrarily convoluted, a naive serial combination of AML with SMT will also fail because most of the AML-generated samples would not have a corresponding packet sequence that is realizable. 
Our second insight to address this challenge is that not all AML-generated features need to be honored for the sample to be adversarial.  We design an optimization loop containing an AML component (e.g., PGD or ZOO), an SMT solver, and the target \netApp, which iteratively checks whether a subset of the AML-produced adversarial features (or perturbations) can result in a realizable packet sequence until
enough adversarial and realizable samples are collected. 
Instead of trying every possible subset of features, our third insight is to identify the most vulnerable adversarial features \ie those that are both easy to tweak based on the threat model and important enough for the target ML model to affect its output. Critically, though, \sys might opt out of constraints that make the generated samples adversarial but not constraints that make samples realizable or semantics preserving, which are always satisfied.

%% file: sections/4-design.tex
\vspace{-0.35cm}
\section{Design}

\subsection{\sys end-to-end view}

We provide an end-to-end view of how \sys generates adversarial packet sequences, shown in Fig.~\ref{fig:e2e}. The procedure includes 8 steps, starting from feeding in the original packet sequence to getting the adversarial and realizable packet sequence returned by \sys.

Step \textcircled{1}: Given an original packet sequence, we use the feature engineering module in the target application to extract its features. 
Step \textcircled{2}: After getting the features, \sys uses the canonical AML methods (PGD, ZOO) to generate an adversarial feature vector (sample). The generation process interacts with the ML classifier to ensure the adversariness of the generated sample.
Note that the generated adversarial features are not strictly required to be within a predefined $\ell_\infty$-ball distance around the original features since the capability of the attacker (threat models) for the networking applications is different from the image domain described in ~\S\ref{subsec:aml-promise}. 
Step \textcircled{3}: Given the generated adversarial feature, \sys runs an iterative process to find (multiple) adversarial and realizable packet sequences using a feature selector and an SMT solver. The feature selector first determines the importance of each feature in keeping the generated features adversarial. 
Step \textcircled{4}: The feature selector begins to construct a selected feature list by considering one new feature each time, starting by appending the most important feature and proceeding in descending importance order. 
Step \textcircled{5}: Given the selected features, we encode these features' dependencies, threat model constraints, and networking constraints into formulas and query the SMT solver to find a packet sequence to satisfy them. 
Step \textcircled{6}: If the SMT solver returns ``UNSAT'', \ie it cannot find a packet sequence, the feature selector pops out the latest appended feature from the selected feature list.  The iterative process continues to step \textcircled{4} to append the next most important feature and repeat step \textcircled{5} and \textcircled{6} until the first $k$ most important features are considered. 
Step \textcircled{7}: The packet sequence(s) found by the SMT solver are guaranteed to be realizable since the threat models are encoded into the satisfiable formulas. 
Finally, we need to confirm the adversariness of these realizable packet sequences. We feed these packet sequences into the feature engineering module again to get the corresponding feature vectors. 
Step \textcircled{8}: We feed these feature vectors into the ML classifier to get the predicted label. We compare the prediction result with the ground truth label of the original packet sequence. If they are different, the realizable packet sequences found by the SMT solver are adversarial, and the feature vectors correspond to realizable packet sequences.
Note that these packet sequences are very likely to be adversarial, given that the feature selector prioritizes the use of the most important features.


\subsection{SMT formulations}


\change{
\sys incorporates an SMT solver for generating adversarial and realizable packet sequences.
Concretely, the SMT solver encapsulates constraints for feature dependencies, the threat model, networking rules, and semantics, determining their satisfiability, i.e., whether a sample satisfying these constraints exists.
} 


\change{To provide a clearer illustration of this encoding, let's consider a concrete example.}
Consider an original network flow that only has three packets, p= \( \{p_1^+, p_2^-, p_3^+\} \) where $p_i^{\{+,-\}}=(\Delta t_i, l_i)^{\{+,-\}}$. ${\{+,-\}}$ represents the direction of the packet, where $+$ means packets sending from the connection initiator to the receiver (forward packet) and $-$ vice versa. $t_i$ and $l_i$ represent the inter-arrival time and packet length for each packet $p_i$. \change{For this example, let us assume that the adversarial features generated by the AML component specify the following: the average inter-arrival time is $k$, the maximum inter-arrival time is $m$, and the average packet length is $j$.}
\change{Assume that the threat model includes delaying and appending dummy payload to forward packets, but to preserve the semantics meaning of the flow, the accumulative delay cannot exceed 20\% of the original duration.}

For this example, we thereby define four variables $\Delta t_1'$, $\Delta t_3'$, $l_1'$ and $l_3'$. They are representing the delay and the appended payload for $p_1^+$ and $p_3^+$. The perturbed flow p' can be formulated as p'$=\{\Bar{p}_1^+, p_2^-, \Bar{p}_3^+\}=\{(\Delta t_1 + \Delta t_1', l_1 + l_1')^+, (\Delta t_2, l_2)^-, (\Delta t_3 + \Delta t_3', l_3 + l_3')^+ \}$. Then, the formula for the SMT solver is
\begin{empheq}[left=\empheqlbrace]{align}
    & (\Delta t_1 + \Delta t_1' + \Delta t_2 + \Delta t_3 + \Delta t_3')/3 = k \label{eq:1}\\
    & (\Delta t_1 + \Delta t_1' = m) \lor (\Delta t_2 = m) \lor (\Delta t_3 + \Delta t_3' = m)  \label{eq:2}\\
    & (\Delta t_1 + \Delta t_1' \leq m) \land (\Delta t_2 \leq m) \land (\Delta t_3 + \Delta t_3' \leq m) \label{eq:3}\\
    & (l_1 + l_1' + l_2 + l_3 + l_3')/3=k \label{eq:4}\\
    & \Delta t_1' \geq 0 \land \Delta t_3' \geq 0 \land l_1' \geq 0 \land l_3' \geq 0 \label{eq:5} \\
    & l_1 + l_1' \leq 1500 \land l_3 + l_3' \leq 1500 \label{eq:6} \\
    & \Delta t_1' + \Delta t_3' \leq 0.2 \times (\Delta t_1 +\Delta t_2 + \Delta t_3)  \label{eq:7} 
\end{empheq}

\change{Equation~\eqref{eq:1}, \eqref{eq:2}, \eqref{eq:3} and \eqref{eq:4} force the generated packet sequences to be adversarial by being consistent with the average (equation~\eqref{eq:1}), maximum (equation~\eqref{eq:2},\eqref{eq:3}) inter-arrival times and the average of packet length (equation~\eqref{eq:4}) that AML specification. Equation~\eqref{eq:5} ensures that the packet sequence complies with the threat model, which includes delaying and appending dummy payload to the forward packets.
Equation~\eqref{eq:6} and ~\eqref{eq:7} formulate the network and semantics constraints.}


\subsection{Adversarial packet sequence identification} 
\label{sec:design:aps_id}

\paragraph{Feature importance determination(line 1 - line 5 in Alg.~\ref{alg:feat_selection}).}
\new{Adhering to all features in the adversarial feature vector that the AML specified could be too computationally expensive for the SMT solver or even infeasible.
Critically, though, not all features are equally important in keeping the packet sequences adversarial.
For instance, given an adversarial vector of two concrete values generated by the AML component, there is often a packet sequence that adheres to only one of them while still being adversarial.
Building on this insight, we design a feature importance ranking algorithm that helps \sys identify feature subsets that play a more important role in the sample's adversariness, thereby streamlining and accelerating the search process.}
Since AML methods are designed to perturb features to maximize the chance that they will be misclassified, the features that have been changed the most play a more important role in adversariness.  
Thus, after normalizing all the features, the importance of each feature is determined by the difference between the adversarial and the original value of that feature. Using this importance ranking, we can prioritize adversarial-feature constraints, which are crucial to keeping the flow adversarial.


\paragraph{Find adversarial and realizable packet sequences using iterative testing (line 19 - line 28 in Alg.~\ref{alg:feat_selection}).}
%
\new{
After quantifying feature importance, \sys leverages this information to maximize the likelihood of quickly identifying adversarial and realizable samples. Rather than starting with the full set of adversarial-features constraints or testing random permutations—both of which would significantly increase computational overhead in the SMT solver—\sys employs the following algorithm to optimize the process.}
%
%

\begin{figure}[ht]
\centering
\begin{subfigure}{.475\linewidth}
    \centering
    \includegraphics[width=.95\linewidth]{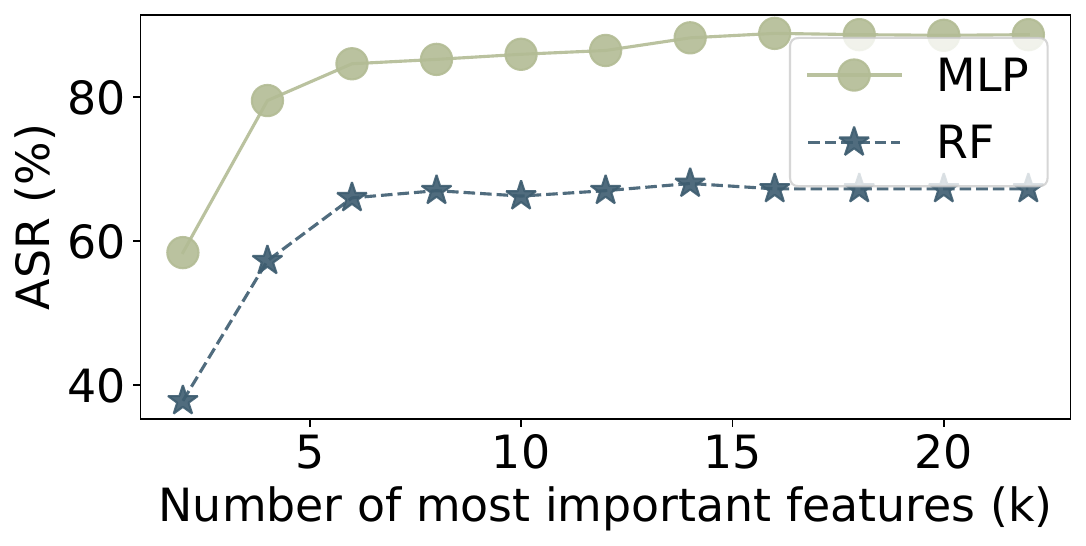}  
    \caption{End-host attacker}
    \label{}
\end{subfigure}
\begin{subfigure}{.475\linewidth}
    \centering
    \includegraphics[width=.95\linewidth]{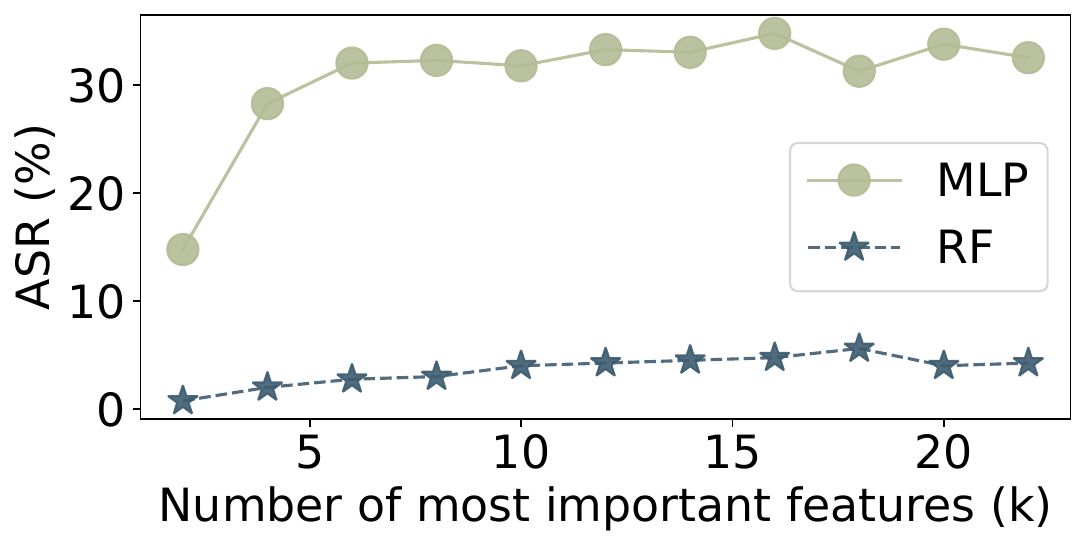}  
    \caption{In-path attacker}
    \label{}
\end{subfigure}
\caption{\new{
The impact of $k$ in ASR (\ie in \sys ability to generate examples) against MLP and RF.  
The ASR increases with $k$, but levels off once 
$k$ reaches a certain threshold.
}
}
\label{fig:important_features}
\end{figure}
Starting with an empty feature list, we iteratively test whether adding the formulas of the next most important (not already explored) feature $f_i$ results in a conforming packet sequence output from the SMT solver. 
If it does, $f_i$ is appended to the selected features list. Otherwise, the feature is not selected.
By adding more and more features, the found packet sequences are more likely to be adversarial. 
 \new{Even after the SMT solver finds a packet sequence that is adversarial, the algorithm continues appending more features until it appends $k$ features, to cover a wider spectrum of adversarial examples. The exact value of $k$ can be set and easily fine-tuned by the user. Concretely, \sys finds more diverse adversarial examples as one increases $k$ (up to a point), as we observe in Fig.~\ref{fig:important_features}, but will also take longer to compute them. In practice, one only needs to avoid setting $k$ too low, making half the number of features a reasonable default.}

\change{Thanks to these two key logics, \sys is able to identify adversarial packet sequences. The algorithm details are depicted in Alg.~\ref{alg:feat_selection}.}

\begin{algorithm}[hbt!]
\caption{\new{find adv pkt sequences with iterative testing}}\label{alg:feat_selection}
\scriptsize
\begin{algorithmic}[1]
\Require original\_sample \Comment{The original and benign packet sequence}
\Require adv\_features \Comment{Adversarial features generated by AML}
\Require k \Comment{Configured number of important features}
\Ensure solutions\_set

\Function{Feature\_Rank}{original\_sample, adv\_features}
\State original\_feature = feature\_engineer(original\_sample)
\State sorted\_feature\_list = sort(abs(adv\_features - original\_features))
\State return sorted\_feature\_list \Comment{Sorted by importance descending order}
\EndFunction
\State \strut
\Function{SMT\_Solver}{selected\_features, original\_sample}
\If{ SMT can find solutions consistent with selected\_features}
    \State return solutions
\Else
    \State return null
\EndIf
\EndFunction
\State \strut
\State sorted\_feature\_list $\gets$ ()
\State selected\_features $\gets$ ()
\State solutions\_set $\gets$ ()
\State sorted\_feature\_list $\gets$ Feature\_Rank(original\_sample, adv\_features)
\For{$i$ in 1:k}
    \State selected\_features.append(sorted\_feature\_list[$i$])
    \State solution $\gets$ SMT\_Solver(selected\_features, original\_sample)

    \If{solution is null}
        \State selected\_features.remove(sorted\_feature\_list[$i$])
    \Else 
        \State solutions\_set.append(solution)
        \State Check adversariness for this solution
    \EndIf
\EndFor
\end{algorithmic}
\end{algorithm}

\subsection{\sys' SMT optimizations}
\label{subsec:smt-opt}
\change{While preferentially enforcing subsets of adversarial features improves scalability, there are scenarios where the SMT solver's limitations become a more difficult bottleneck. For example, there will be cases in which finding an adversarial example requires adhering to all feature constraints or cases where the most significant feature constraint is so complex that the solver struggles.}
The complexity depends on many factors, such as the length of the packet sequence, the formulation of capabilities, the number of features to be solved, and the difficulty of solving each feature.
In this section, we focus on optimizing the packet sequence length and the formulation of capabilities.


\myitem{Network flow chunking and parallelization.}
A longer packet sequence requires more variables to formulate, which implies a much larger search space for the SMT solver.
We address this problem with a combination of a timeout mechanism, network flow chunking, and parallelization.

We first set a time budget for the SMT solver.
If the SMT solver cannot find packet sequences within a timeout window, we decrease the computation complexity by dividing a flow into several chunks evenly. The SMT solver is used to find the packet sub-sequences for each chunk. Since each chunk has fewer packets than the whole flow, it is more likely that the SMT solver will be able to find solutions within the timeout limits. 
After splitting a flow into multiple chunks, some features are difficult to formulate strictly. For example, when working on a complete flow, it is easy to formulate the maximum packet length feature. After splitting the flow into multiple chunks, the maximum packet length can appear in any chunk. Strictly formulating this feature would require a complicated algorithm, which is in conflict with our motivation to decrease computation complexity and is unnecessary for our use case. 
To solve this problem, we consider three different types of features.
For the first type of features, satisfying their formulas for each sub-sequence will likely result in satisfied formulas for the whole sequence. 
For example, such features are \{max, mean, median, min, std\} of \{interarrival times, packet length\} and number of \{packets, bytes\} per second. For these features, SMT solves their formulas per chunk. 
For the second feature type, the accumulation of each chunk's feature values is equal to that of the whole flow. One typical example is the total duration. 
For these features, the SMT solver is finding sub-sequences in each chunk consistent with $\frac{feature}{\#chunks}$. 
For other features \change{which exhibit convoluted correlations between the sub-sequences and the whole sequence},
such as the number of unique packet lengths, we are not aiming to be consistent with the whole-flow formulas when finding the sub-sequences for each chunk.
 \new{ The number of chunks has minimal impact on ASR. Indeed, we found that even doubling their number (from 10 to 20 chunks) changes ASR by 3\%. That is because once the solver is able to find sub-sequences for each chunk, continuing to split into smaller chunks does not improve ASR.}

Finally, \sys' use of the solver is highly parallelizable, facilitating further speed ups. Indeed, the solver for each original input, chunk, and even set of adversarial features of a given instance can run independently.


\myitem{Inefficient formulation of capabilities can be simplified.}
Unfortunately, not all the capabilities can be formulated efficiently. 
The capability to inject packets at any position is one of them. The underlying reason is that multiple dummy packets can be injected at any position. Each combination requires an invocation of the SMT solver. For example, given a flow with four packets, $f = [p_1^+, p_2^-, p_3^+, p_4^-]$, injecting packet $p_5^{*+}$ by $[p_1^+, p_2^-, p_3^+, p_4^-, p_5^{*+}]$ or $[p_1^+, p_2^-, p_3^+, p_5^{*+}, p_4^-]$ can cause different solutions on $p_5^{*+}$ when asking the packet sequence to comply with the features of backward flow interarrival times.
Suppose a threat model that allows injecting at most $j$ packets and the packet sequence length is k; formulating this capability into formulas requires SMT to run $\sum_{m=0}^{j}\binom{m+k}{m}$ times to cover all combinations. 
%
Therefore, packet injection at any position is very inefficient, and we consider simplifying the formulation to increase the generation speed.
Although packet injection at any position requires the SMT solver to find packet sequences multiple times, injecting packets at the end of the flow is efficient since there is only one injection position. Thereby, the SMT solver only needs to run once.
We leverage the flow chunking and simplify the implementation of any-position injection by injecting only at the end of each chunk. As we increase the number of chunks, the number of packets per chunk is decreased and the final effect is closer to packet injection at any place.

Fig.~\ref{fig:insight2-speed} shows the sample generation speed after the discussed optimizations. \new{\sys generates \textasciitilde 1.7 samples/sec, which is efficient, considering \NetApps are rarely re-trained.}
%


\subsection{Iterative Adversarial Training with \sys}
\label{subsec:at}
Adversarial training is often used to enhance the robustness of ML models.
During authentic adversarial training, an AML method (\eg PGD) replaces each benign sample with an adversarial one for the insufficiently trained version of the model. While useful, this method can hurt the accuracy of an ML model~\cite{grabinski2022robust,salman2020adversarially} and only applies to NN-based models. 
To avoid these shortcomings, we design an alternative way of adversarial training, in which we augment the training dataset with adversarial samples generated for the sufficiently trained model to fine-tune it. This methodology is model-agnostic, meaning that it can work with any processing pipeline. 
We also find in \S\ref{subsec:kf} that this training methodology improves robustness without hurting the model's accuracy.

\change{For robustification, \sys generates multiple adversarial samples for each benign counterpart in total to cover diverse adversarial samples.
Using \sys, we generate around 1.16-180 samples for each benign sample during robustification based on different threat models, applications and ML models.
Along with adversarial samples, we include non-adversarial ones generated by \sys. This is beneficial for training because the non-adversarial samples are closer to the decision boundary compared to their benign counterpart, despite being correctly classified. Including those samples can help form a more refined decision boundary.}

%% file: sections/6-eval.tex
\vspace{-0.35cm}
\section{Evaluation}
\vspace{-0.15cm}
\label{sec:eval}

In this section, we evaluate \sys in two ways.
\change{First, we evaluate \sys' capability in generating adversarial realizable and semantic-preserving flows and compare it with two baselines, Amoeba~\cite{amoeba} and BAP~\cite{BAP}. 
We find that \sys is 70\% more likely to find such
flows compared to Amoeba in median and 2x more
likely compared to BAP. 
Second, we evaluate the effectiveness of \sys-generated samples in improving the robustness of the targeted \netApps. 
We find that \sys is 142\% more effective in improving robustness compared to Amoeba without sacrificing the model accuracy. 
Interestingly, we empirically find \sys enhances the model's robustness even against attackers that have different (and even strictly more) capabilities than what was assumed during sample generation.}
We begin by introducing the evaluation setup before discussing some key findings.


\vspace{-0.25cm} 
\subsection{Setup}
\label{subsec:setup}

\mypara{Applications \& Datasets}
We evaluate \sys in three \netApps with diverse objectives, flow statistical patterns, and classification granularity, providing a representative sample of networking applications. \new{We evaluate several implementations of each \netApp, including the exact implementation described in the paper from which each dataset is sourced.}

\change{
    \textbf{VPN:} VPN traffic detection aims at identifying flows that are VPN among encrypted traffic flows\cite{iscxvpn2016,parchekani2020classification,VPN2020sequence,VPN2022TPIPD}. ISCXVPN2016~\cite{iscxvpn2016} is one of the commonly used labeled datasets and includes both VPN and non-VPN traffic. The dataset contains more than $20$ million packets grouped into $8,577$ bi-directional flows. Of those, $71$\% are non-VPN, and the rest are VPN flows. 
    
    \textbf{APP:} Application identification aims at predicting the application that given traffic serves and is useful for network management tasks such as anomaly detection and resource allocation~\cite{TC2005,utmobile,TC2017lopez,Velan2015ASO,Wang2019survey}. 
     UTMobileNetTraffic2021~\cite{utmobile} is a commonly used dataset that includes mobile traffic from 18 popular applications (\eg Dropbox, Google Drive, Facebook).
    The selected dataset contains 7,134 bi-directional flows. 
    
    \textbf{QoE:} Quality of Experience inference aims at inferring QoE metrics like resolution and video bitrate from packet traces\cite{sharma2023vca,Carofiglio2021Cha,Nikravesh2016QoE,Yan2015EnablingQL,Hossain2017quality}. QoE inference is useful for stakeholders, such as ISPs and service providers, due to the rise in the use of Video Conferencing Applications (VCAs). 
    VCAML~\cite{sharma2023vca} is the dataset that contains a large number of single-directional video conferencing flows in a per-second granularity from Google Meet, Microsoft Teams, and  Cisco Webex.  The total number of samples is 37,274, which includes 11 different resolutions, such as 720p, 360p.
}

\change{
The features extracted by the feature engineering module for different applications are listed in the Appendix in~\S\ref{apdx:sec:feat}. 
}

\mypara{Application implementations}
\label{hp-explain}
For each application, we 
train four commonly used ML classifiers independently. They are Multilayer Perceptron (MLP), Random Forest (RF), Transformer (TF) and Convolutional Neural-Network (CNN) \change{to cover multiple ML approaches, and processing pipelines (\ie end-to-end differentiable or not).}
The details of the hyperparameters used for each model are explained in the Appendix in \S\ref{apdx:sec:hp}. Table~\ref{tab:basic_ml_perf} shows the performance of these three \netApps. For each dataset, the training set ($D_{train}$) and the testing set ($D_{test}$) are following a 80\%:20\% random split. ML models are trained on $D_{train}$, and the performance is evaluated on $D_{test}$.

\begin{table}[h]
\centering
\small
\begin{tabular}{c|c|c}
\hline
Application & \begin{tabular}[c]{@{}c@{}}Accuracy\\ (Vanilla)\end{tabular} & \begin{tabular}[c]{@{}c@{}}Accuracy\\ (In-path/End-host robustified)\end{tabular} \\ \hline
APP & {[}0.7666, 0.8206{]} & {[}0.7400, 0.8136{]} \\ \hline
VPN & {[}0.8902, 0.9400{]} & {[}0.8596, 0.9375{]} \\ \hline
QOE & {[}0.7819, 0.8221{]} & {[}0.7898, 0.8173{]} \\ \hline
\end{tabular}
\caption{\change{Performance of vanilla ML classifiers for each application. Our trained models demonstrate reasonably good accuracy for each application. After robustification via \sys, the robustified ones' accuracy doesn't drop.
}}
\label{tab:basic_ml_perf}
\vspace{-0.4cm}
\end{table}

\begin{figure*}[h]
\centering
\begin{subfigure}{.8\textwidth}
    \centering
    \includegraphics[width=.45\linewidth]{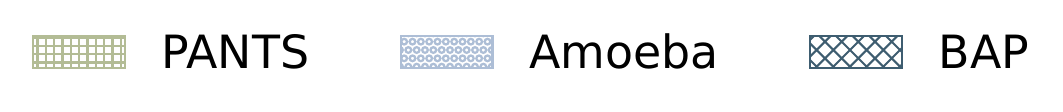}  
\end{subfigure}
\begin{subfigure}{.33\textwidth}
    \centering
    \includegraphics[width=.95\linewidth]{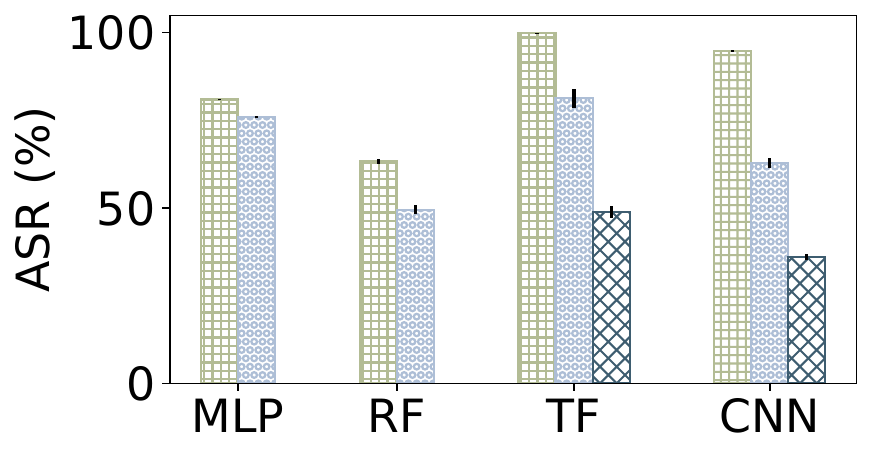}  
    \caption{APP End-host attacker}
    \label{app-strong-asr}
\end{subfigure}
\begin{subfigure}{.33\textwidth}
    \centering
    \includegraphics[width=.95\linewidth]{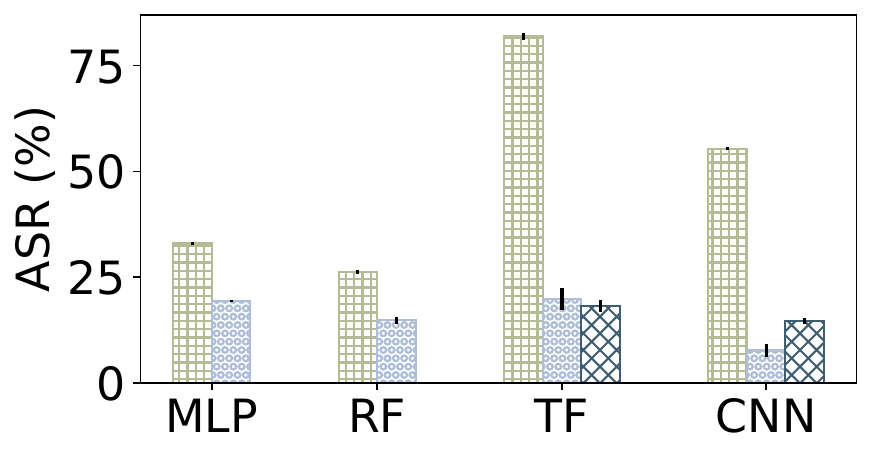}  
    \caption{VPN End-host attacker}
    \label{vpn-strong}
\end{subfigure}
\begin{subfigure}{.33\textwidth}
    \centering
    \includegraphics[width=.95\linewidth]{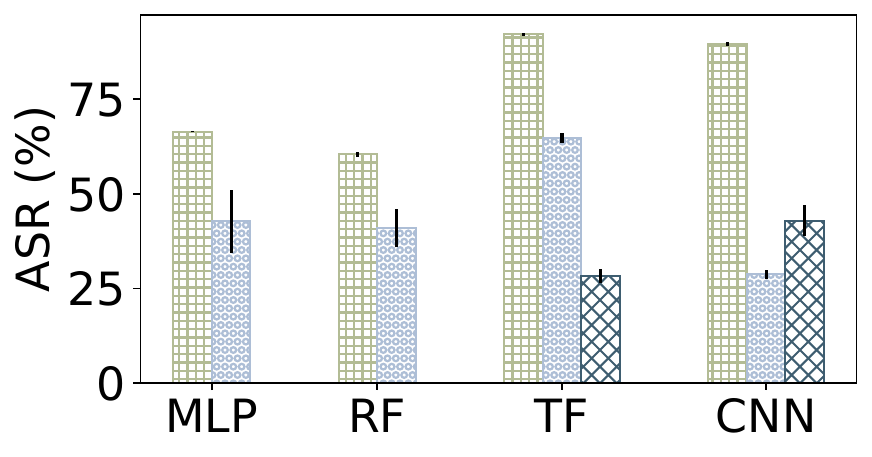}  
    \caption{QOE End-host attacker}
    \label{qoe-strong}
\end{subfigure}
\begin{subfigure}{.33\textwidth}
    \centering
    \includegraphics[width=.95\linewidth]{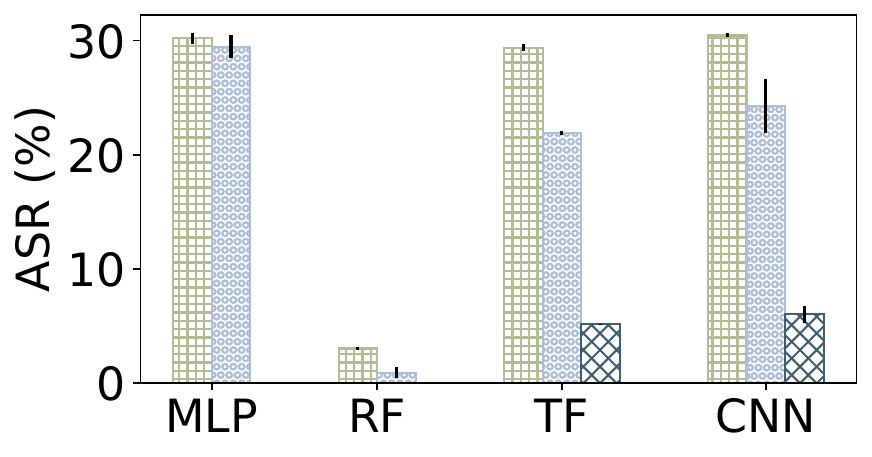}  
    \caption{APP In-path attacker}
    \label{app-weak}
\end{subfigure}
\begin{subfigure}{.33\textwidth}
    \centering
    \includegraphics[width=.95\linewidth]{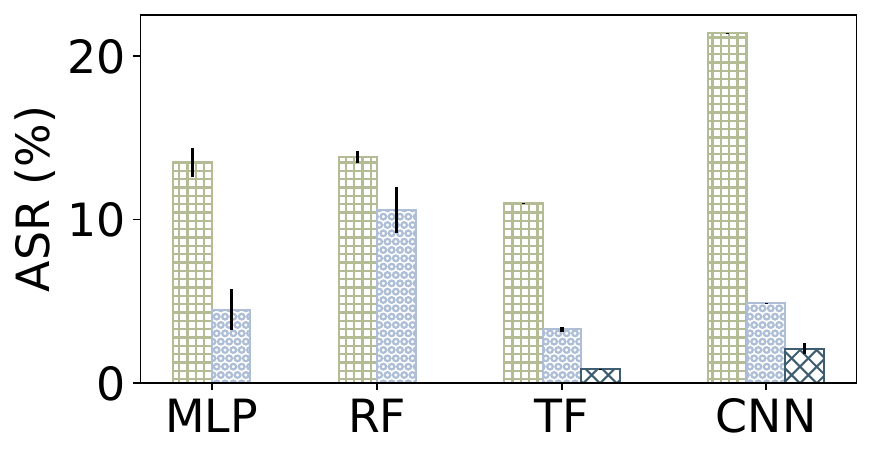}  
    \caption{VPN In-path attacker}
    \label{vpn-weak}
\end{subfigure}
\begin{subfigure}{.33\textwidth}
    \centering
    \includegraphics[width=.95\linewidth]{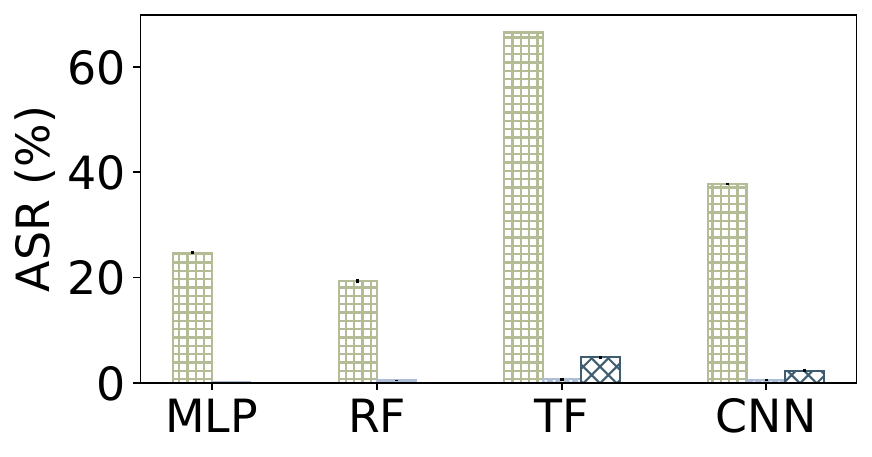}  
    \caption{QOE In-path attacker}
    \label{qoe-weak}
\end{subfigure}
\caption{The attack success rate (ASR) of \sys, Amoeba and BAP for various ML models, applications and threat models. \sys has a much higher ASR compared to Amoeba and BAP, demonstrating its ability to effectively generate adversarial samples, which can be used for debugging, fine-tuning, and robustness assessment.}
\label{val1}
\vspace{-0.1cm}
\end{figure*}

\mypara{Attacker's capability}
We evaluate two threat models that correspond to attackers located at the end host and in-path. 
The attackers at the end host (\ie server or client) can delay packets, inject packets, and append dummy payload to the existing packets of one direction. 
To preserve the flow's semantics, we consider that the end-host attacker can only delay the flow up to 20\% of the original flow's duration, inject at most 20 packets, and append dummy payload to at most 20\% of the packets. 
The in-path attackers can only delay packets of one direction at most 20\% of the original flow's duration.
\new{ We selected these two models because they represent realistic and rational attackers, but thanks to \sys flexibility
one can easily add more threat models  by adding the relevant constraints as we do in Fig.~\ref{fig:more-threat}.}


\mypara{Baselines}
Since there is no related framework that aims to help the operator assess and enhance the robustness of \netApps, we compare \sys against the SOTA adversarial generation methods, Amoeba and BAP \new{and a synthetic data generator, NetShare}. 

\textbf{Amoeba}~\cite{amoeba} is an RL-based proposal to attack ML-based censorship. Although designed for a different goal, Amoeba also generates adversarial samples against applications that are end-to-end differentiable and that include a non-differential or non-invertible module. 
Amoeba supports packet delay, dummy payload appending, and packet split into multiple ones. For a fair comparison, we extend Amoeba's capability to comply with the two evaluated threat models. We also train Amoeba's agent to respect the flow's semantics. We explain the details of our extension in the Appendix in \S\ref{apdx:sec:amoebaext}.

\textbf{BAP}~\cite{BAP} is to train a generator to create perturbations for each packet. Since BAP does not support applications that include non-differential and non-invertible components, we only consider it for TF and CNN (\ie not with MLP and RF). The original implementation of BAP can only support perturbing every packet, so we extend BAP by adding a mask during training and inference to ensure reliability (\eg only act on forward packets). Although BAP supports per-packet constraints, it does not support flow-wide constraints such as those needed for semantics preservation. 

\new{\textbf{NetShare}~\cite{netshare-sigcomm2022} is a GAN-based synthetic data generator specialized in producing high-fidelity network packet traces. After NetShare is trained on raw packet sequences, it can generate synthetic ones that closely match the distribution of the original data. NetShare can successfully match the convoluted data distribution from backbone traces such as CAIDA~\cite{CAIDA} and data center traces such as UNI1~\cite{benson2010network}.}

\mypara{Success Metric} 
\new{To quantify the robustness of \netApps, we report the Attack Success Rate (ASR), defined as the ratio of realizable, semantic-preserving adversarial samples to the total original samples for each technique, following standard practice~\cite{carlini2017towards}. 
ASR provides a meaningful measure of robustness, as it directly evaluates a system's resilience under stress and adversarial conditions. We also report accuracy for completeness and to assess performance under typical conditions. \final{Accuracy refers to the proportion of successful classifications on the \textbf{original} test set (not adversarial), which is drawn from the same data source (hence the same distribution) as the training set.\footnote{High ASR is orthogonal to high accuracy. \netApps can both have high accuracy and ASR, meaning that they perform well under standard conditions (high accuracy), but are susceptible to attacks (high ASR).}}
}

\mypara{Testbed} Appendix \S\ref{apdx:sec:testbed} 
 details the used CPU/GPU machines.

\vspace{-0.2cm}
\subsection{Key Findings}
\label{subsec:kf}
\label{subsec:findings}

\begin{finding}
    {\sys is 70\% more likely to find adversarial samples compared to Amoeba in median and 2x more likely compared to BAP.}
\end{finding}



We compare \sys with Amoeba and BAP in their ability to generate adversarial flows under two threat models against three \netApps implemented with four models and report the ASR.
As we observe in Fig.~\ref{val1}, \sys clearly outperforms baselines in all cases. More specifically, \sys achieves a 35.31\% ASR (median across all cases), whereas Amoeba and BAP achieve 19.57\% and 10.31\%, respectively.  
BAP's inefficiency stems from its inability to incorporate semantic constraints, \eg flow-level constraints. 
 \change{Amoeba outperforms BAP (despite being a black-box approach), but \sys is still 51\% more likely than Amoeba to identify adversarial samples under the end-host threat model (median) and 2x more likely under the in-path threat model. This highlights that while Amoeba is effective at uncovering adversarial examples in scenarios with abundant opportunities (i.e., for stronger attackers), it struggles when such examples are less apparent.}


Further, \sys's ASR is more consistent across runs in all the test cases. This is important as an operator is unlikely to trust the assessment of a tool that reports a different ASR in every run \ie varying levels of vulnerability for a given \netApp.
Concretely, the average standard deviation across all the test cases is 0.77 for \sys, while for Amoeba  3.12 and for BAP 2.61. 
For some cases, specifically, Amoeba and BAP show much higher fluctuations in ASR compared to \sys. For example,  Amoeba's ASR on attacking TF ranges from 76.15\% to 93.45\% and BAP is from 43.89\% to 54.25\% while \sys' ASR only ranges from 99.75\% to 100.0\% as we observe in Fig.~~\ref{app-strong-asr}. 
The large variance in Amoeba's results stems from a well-known problem in RL: its inherent instability of training. \change{BAP also struggles with instability when training the perturbation generator but is better than Amoeba. 
For \sys, randomness is limited only stemming from \emph{(i)} PGD restarts used in MLP and RF~\cite{madry2017pgd}; and \emph{(ii)} CPU fluctuations affecting the SMT solver's computation speed.} 

\begin{finding}
    {Iterative augmentation with adversarial, realizable, and semantics-preserving samples improves the robustness of an \netApp without hurting its accuracy.}
\end{finding}

\begin{figure}[t]
\centering
\begin{subfigure}{.99\linewidth}
    \centering
    \includegraphics[width=.99\linewidth]{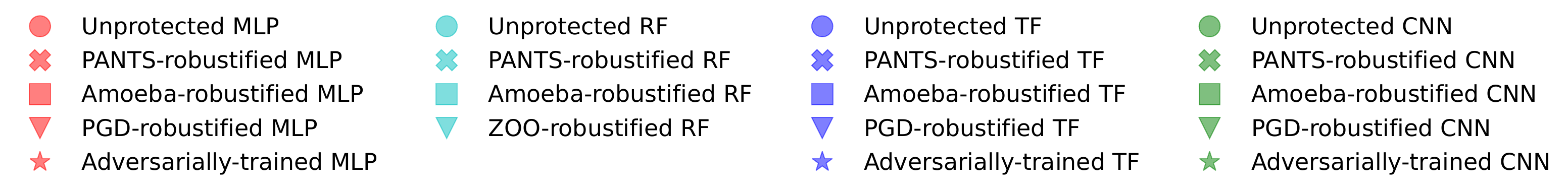}  
\end{subfigure}
\begin{subfigure}{.49\linewidth}
    \centering
    \includegraphics[width=.99\linewidth]{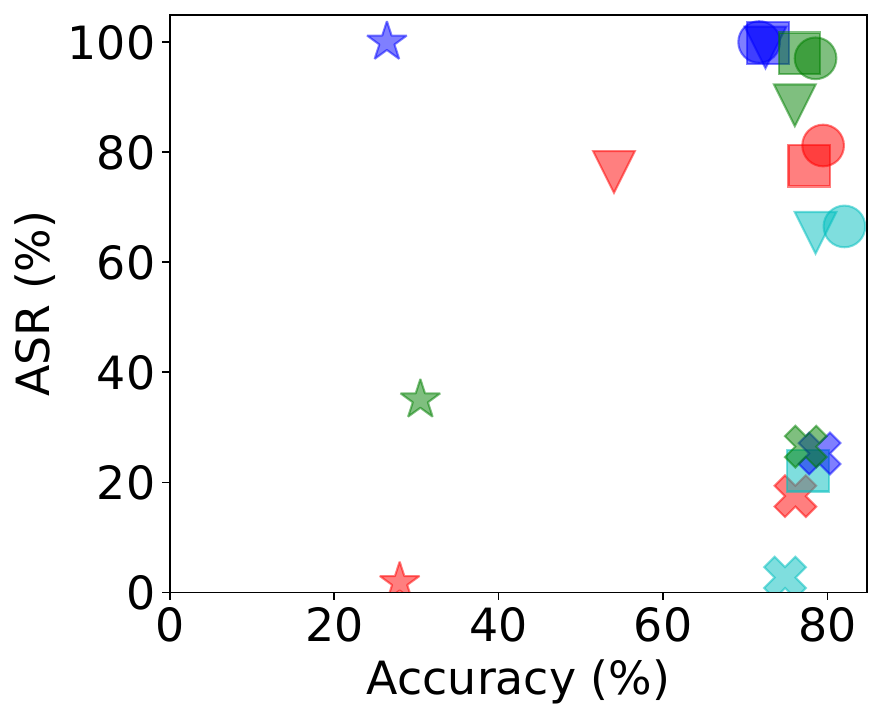}  
    \caption{Attacked by PANTS}
    \label{fig:robust_ml_acc_asra}
\end{subfigure}
\begin{subfigure}{.49\linewidth}
    \centering
    \includegraphics[width=.99\linewidth]{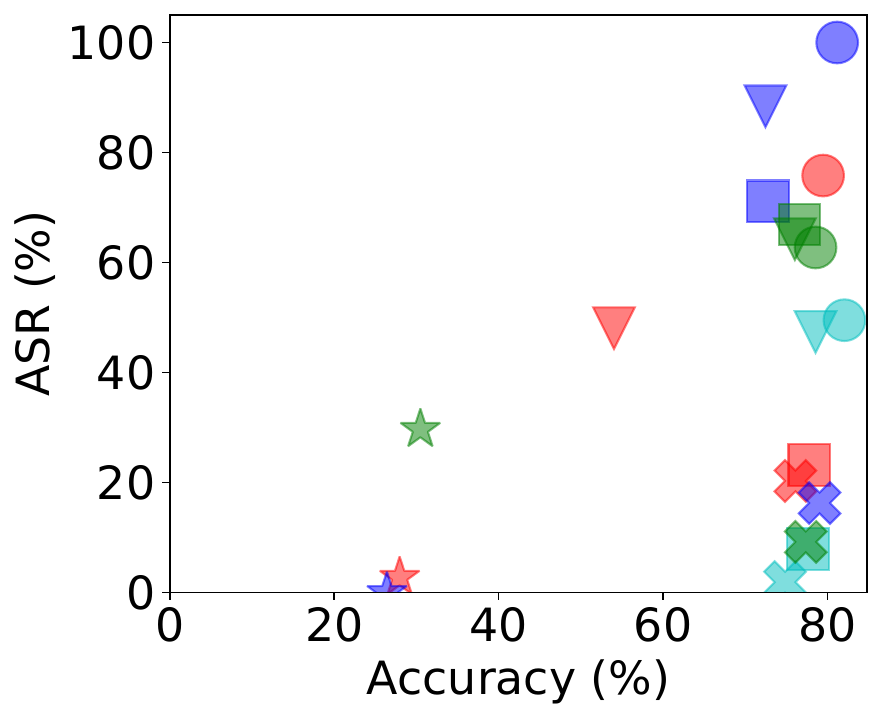}  
    \caption{Attacked by Amoeba}
    \label{fig:robust_ml_acc_asrb}
\end{subfigure}

\caption{The accuracy and ASR for the vanilla and robustified models using different ways of robustification. PANTS-robustified models are robust against both PANTS and Amoeba without sacrificing model accuracy. }
\label{fig:robust_ml_acc_asr}
\vspace{-0.3cm}
\end{figure}
\begin{figure}[t]
\centering
\begin{subfigure}{.45\linewidth}
    \centering
    \includegraphics[width=.95\linewidth]{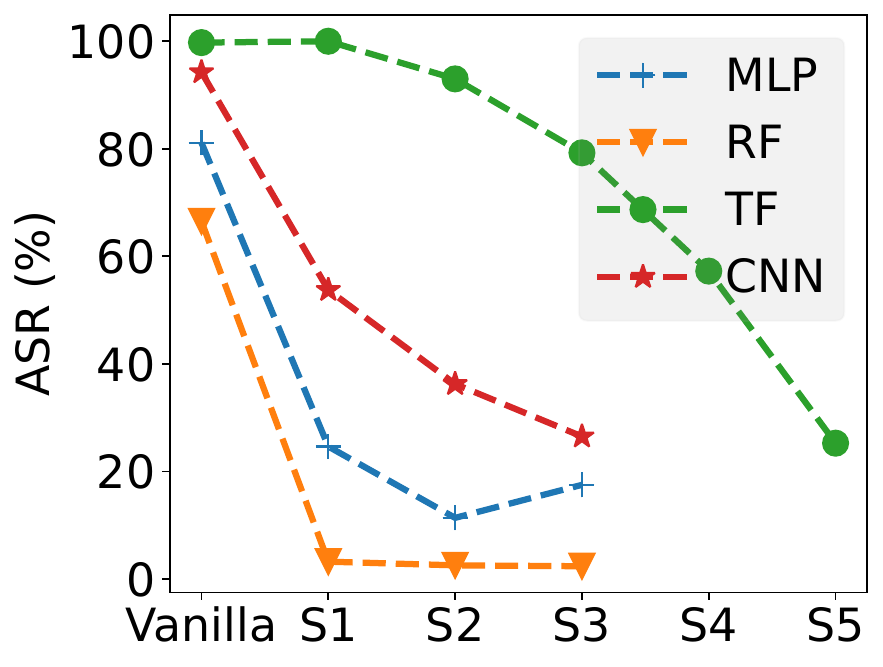}  
    \caption{Robustified for defending end-host attackers.}
    \label{app-strong-asr-drop}
\end{subfigure}
\hspace{10pt}
\begin{subfigure}{.45\linewidth}
    \centering
    \includegraphics[width=.95\linewidth]{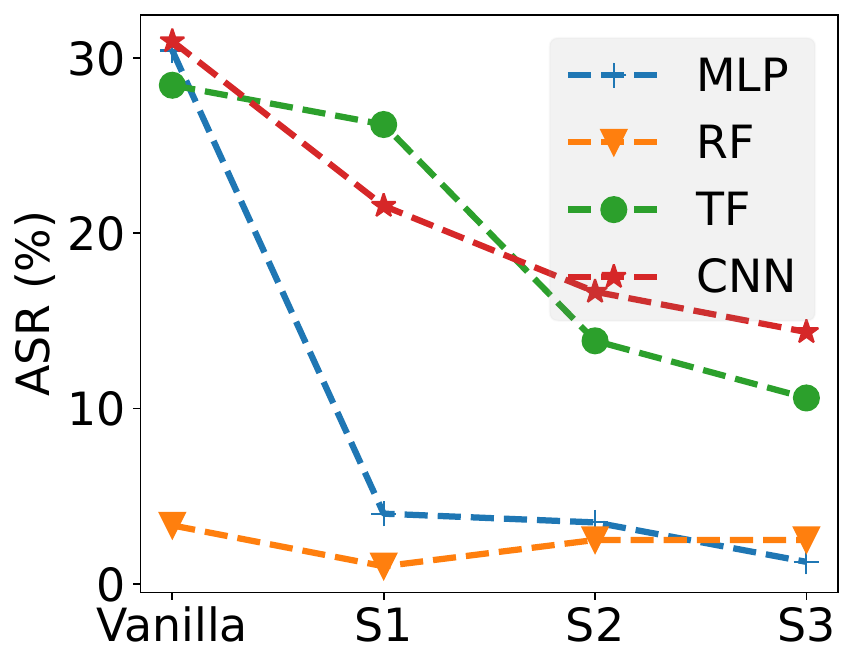}  
    \caption{Robustified for defending in-path attackers.}
    \label{app-weak-asr-drop}
\end{subfigure}
\caption{ASR  calculated by \sys while the \netApp is trained for the i-th time leveraging a dataset augmented with adversarial samples produced by \sys. \netApps are getting more robust against adversarial inputs over time. \new{
For the results using Amoeba's samples for augmentation, please refer to Fig.~\ref{fig:robust_amoeba_asr_drop} in the Appendix.}}
\label{fig:robust_ml_asr_drop}
\end{figure}

\change{To evaluate our iterative adversarial, we compare four approaches for robustifying \netApps. First, we include authentic adversarial training, which integrates PGD into the training process. Next, we include three versions of our iterative augmentation explained in \S\ref{subsec:at}, where adversarial examples are generated by PGD, \sys, and Amoeba. 
Examples generated by \sys and Amoeba are also semantic-preserving.
Fig.~\ref{fig:robust_ml_acc_asr} shows the accuracy and ASR of the robustfied ML models against end-host attackers under the application APP. 
Authentic adversarial training enhances the robustness of the models against attacks from PANTS (Fig.~\ref{fig:robust_ml_acc_asra}) and Amoeba Fig.~\ref{fig:robust_ml_acc_asrb}, but hurts their accuracy.
In contrast, iterative augmentation (our approach) maintains the models' high accuracy (as shown in Table ~\ref{tab:basic_ml_perf}) and could improve robustness subject to the quality of the generated samples.
Concretely, when samples are generated by PGD, iterative augmentation does not help in robustifying the targeted models.
When samples are generated by Amoeba, iterative augmentation robustifies models against samples generated by Amoeba.
Finally, when samples are generated by \sys (\ie as in the complete \sys framework), iterative augmentation 
makes models robust against both PANTS and Amoeba.
Fig.~\ref{fig:robust_ml_asr_drop} and Fig.~\ref{fig:robust_amoeba_asr_drop} offer a more detailed view of this result by plotting the improvement of ASR as a \netApp is under the iterative robustification process. clearly, ASR decreases more dramatically when samples are generated by \sys compared to Amoeba.} 

\begin{finding}
    {\sys improves the robustness tested \netApps 142\% more than Amoeba \new{and \textasciitilde 40X more than NetShare.}}
\end{finding}

\begin{figure}[h]
\centering
\begin{subfigure}{.475\linewidth}
    \centering
    \includegraphics[width=.95\linewidth]{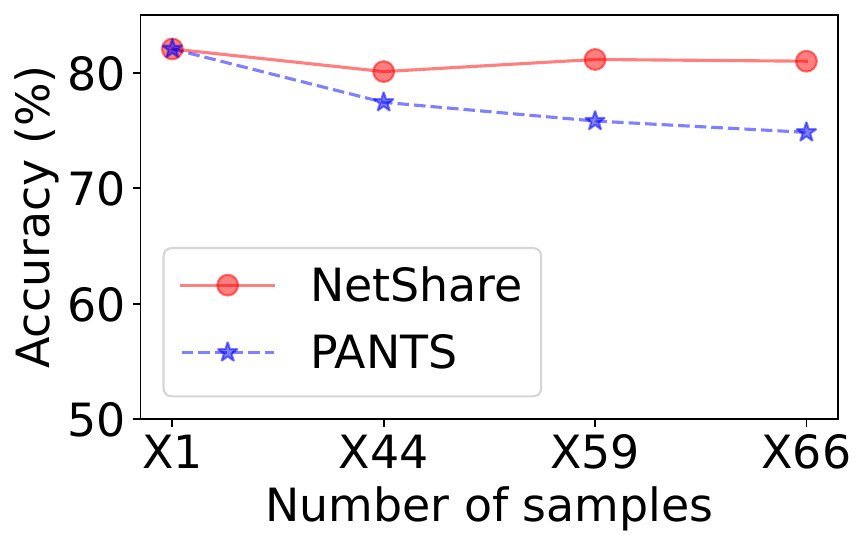}  
    \caption{Accuracy of RF}
    \label{}
\end{subfigure}
\begin{subfigure}{.475\linewidth}
    \centering
    \includegraphics[width=.95\linewidth]{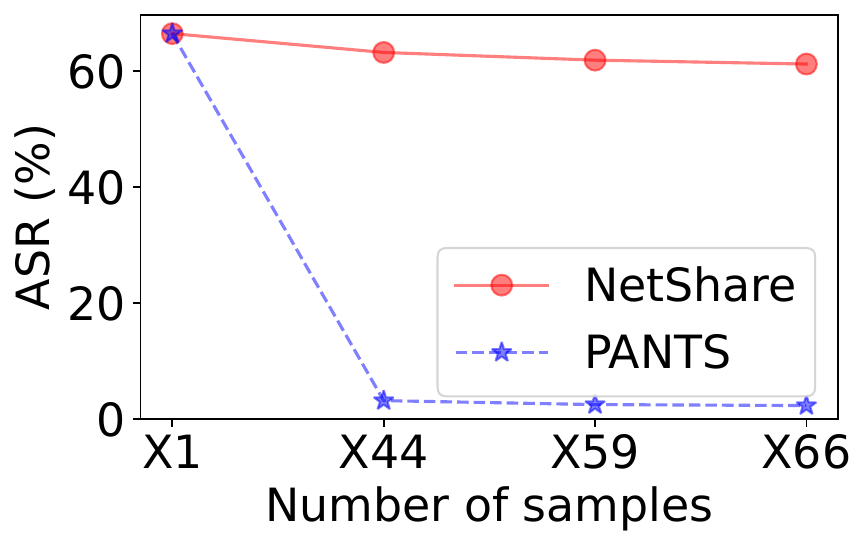}  
    \caption{Vulnerability of RF}
    \label{}
\end{subfigure}
\begin{subfigure}{.475\linewidth}
    \centering
    \includegraphics[width=.95\linewidth]{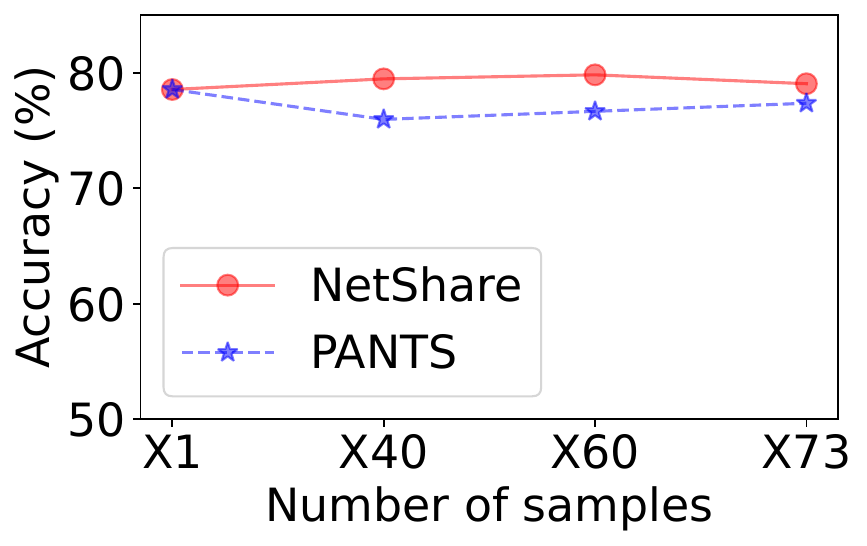}  
    \caption{Accuracy of CNN}
    \label{}
\end{subfigure}
\begin{subfigure}{.475\linewidth}
    \centering
    \includegraphics[width=.95\linewidth]{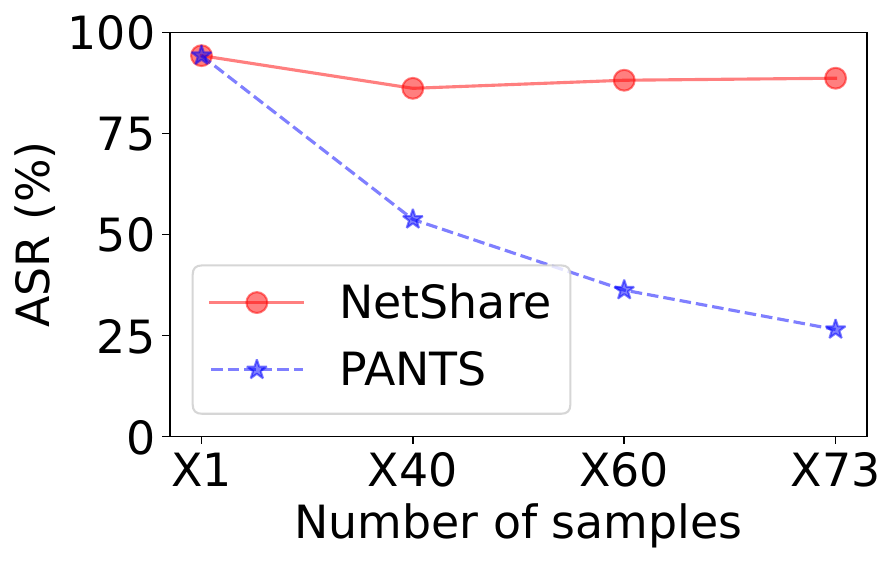}  
    \caption{Vulnerability of CNN}
    \label{}
\end{subfigure}
\caption{\new{
 Accuracy and ASR when training two APP implementations with synthetic data generated by NetShare and adversarial data by \sys. NetShare cannot improve the robustness of \netApps. Similar results for the other two implementations are reported in Fig.~\ref{fig:apdx:netshare_robustification} in the Appendix.
}
}

\label{fig:netshare_robustification}
\vspace{-0.2cm}
\end{figure}


\begin{figure*}[h]
\centering
\begin{subfigure}{.8\textwidth}
    \centering
    \includegraphics[width=.85\linewidth]{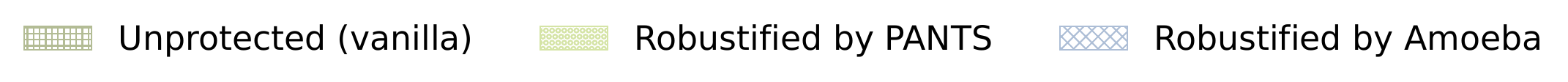}  
\end{subfigure}
\begin{subfigure}{.33\textwidth}
    \centering
    \includegraphics[width=.95\linewidth]{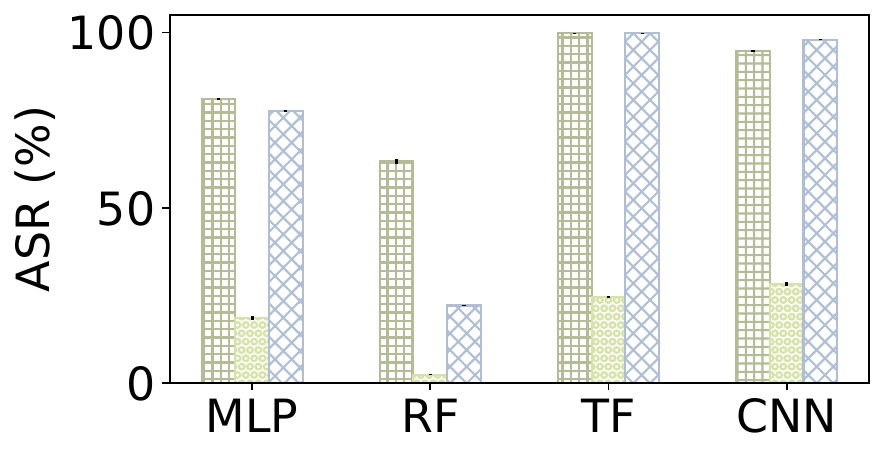}  
    \caption{APP Attacked by \sys}
    \label{}
\end{subfigure}
\begin{subfigure}{.33\textwidth}
    \centering
    \includegraphics[width=.95\linewidth]{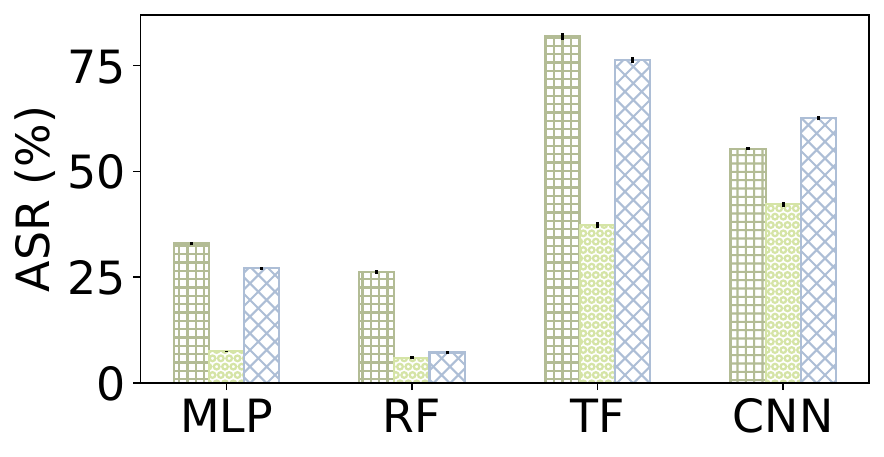}  
    \caption{VPN Attacked by \sys}
    \label{}
\end{subfigure}
\begin{subfigure}{.33\textwidth}
    \centering
    \includegraphics[width=.95\linewidth]{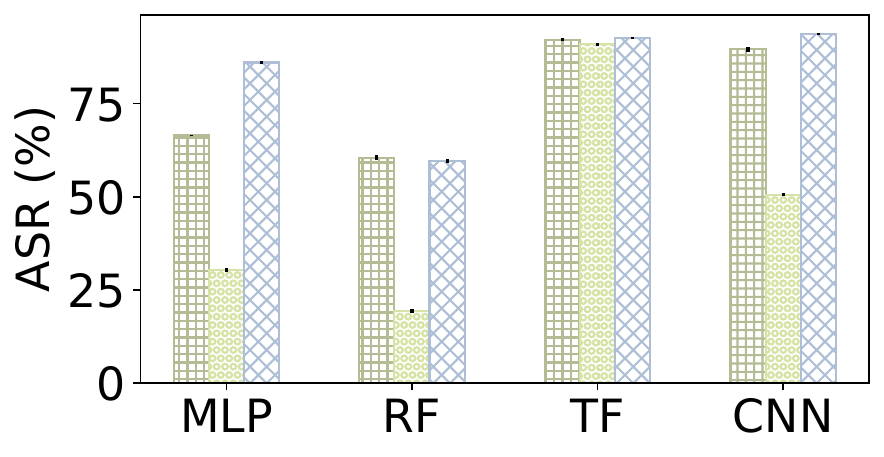}  
    \caption{QOE Attacked by \sys}
    \label{}
\end{subfigure}
\begin{subfigure}{.33\textwidth}
    \centering
    \includegraphics[width=.95\linewidth]{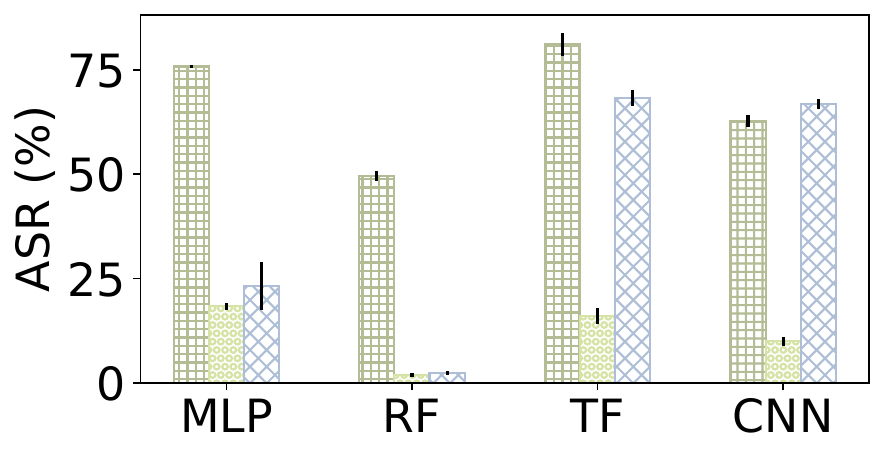}  
    \caption{APP Attacked by Amoeba}
    \label{}
\end{subfigure}
\begin{subfigure}{.33\textwidth}
    \centering
    \includegraphics[width=.95\linewidth]{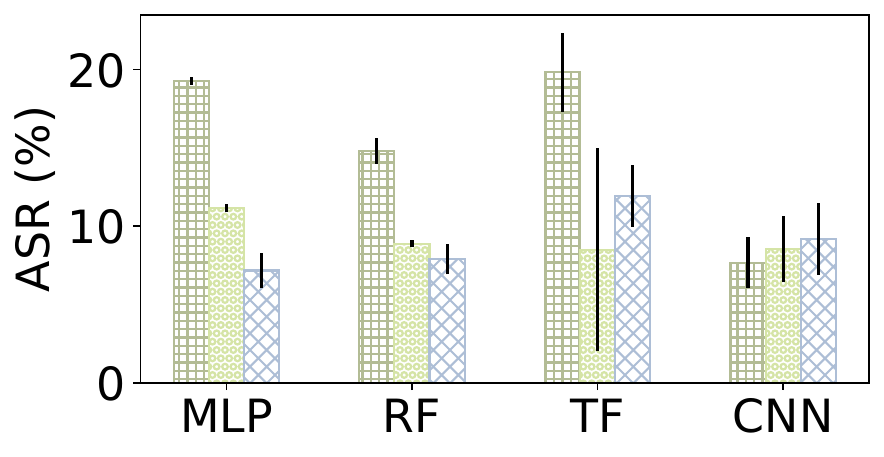}  
    \caption{VPN Attacked by Amoeba}
    \label{}
\end{subfigure}
\begin{subfigure}{.33\textwidth}
    \centering
    \includegraphics[width=.95\linewidth]{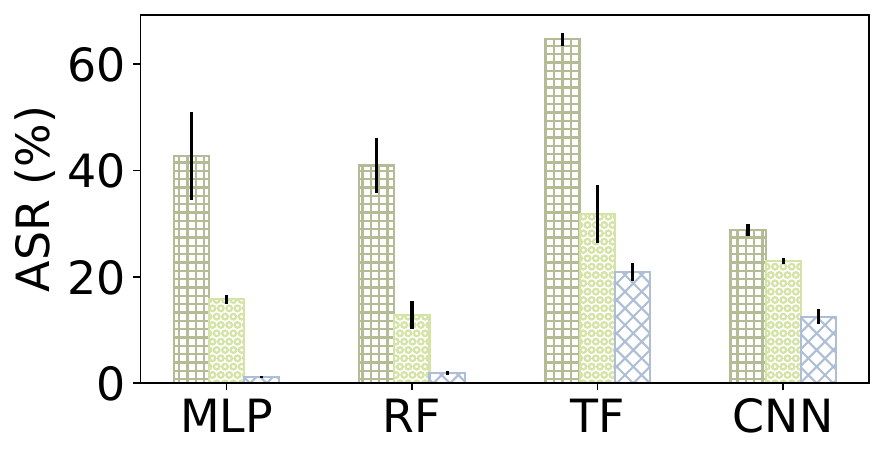}  
    \caption{QOE Attacked by Amoeba}
    \label{}
\end{subfigure}
\begin{subfigure}{.22\textwidth}
    \centering
    \includegraphics[width=.95\linewidth]{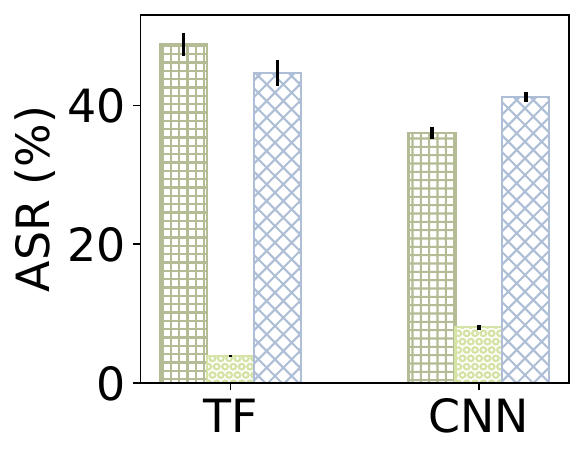}  
    \caption{APP Attacked by BAP}
    \label{}
\end{subfigure}
\begin{subfigure}{.22\textwidth}
    \centering
    \includegraphics[width=.95\linewidth]{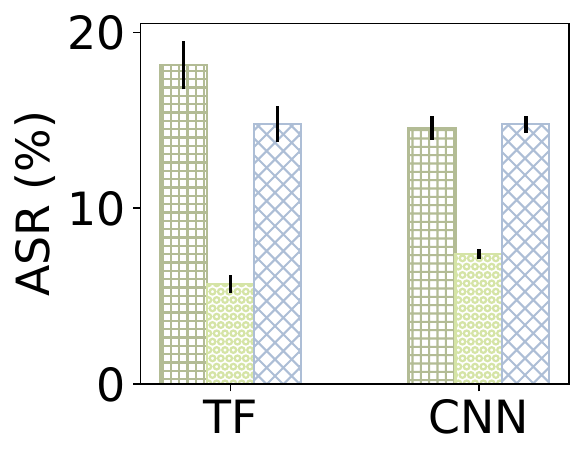}
    \caption{VPN Attacked by BAP}
    \label{}
\end{subfigure}
\begin{subfigure}{.22\textwidth}
    \centering
    \includegraphics[width=.95\linewidth]{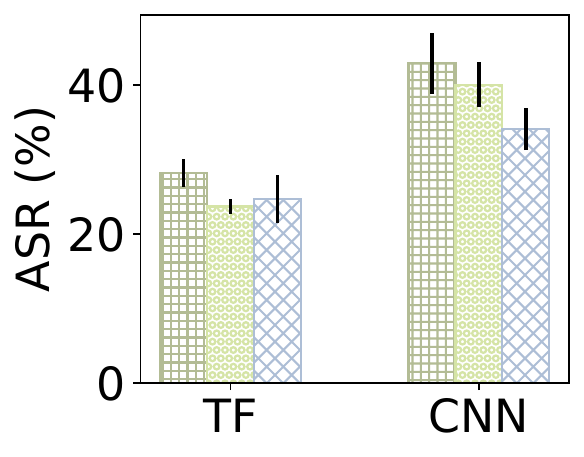}  
    \caption{QOE Attacked by BAP}
    \label{}
\end{subfigure}
\caption{ASR for the vanilla, \sys-robustified and Amoeba-robustified models for the end-host threat model. \sys-robustified's ASR is the lowest. \new{Similar results for in-path threat model are reported in Fig.~\ref{apdx:asr-robust} in the Appendix.}}
\label{fig:val3}
\vspace{-0.4cm}
\end{figure*}

\begin{table}[h]
\centering
\begin{tabular}{l|rrr}
\diagbox[dir=NW,width=2.55cm,height=1.18cm]{\small{Robusified by}}{\strut \small{Attacked} \\ \small{by}} & PANTS & Amoeba & BAP \\ \hline
PANTS & 57.65\% & 49.98\% &  50.52\%\\
Amoeba &  20.89\% & 36.89\%&  7.50\%\\
\end{tabular}
\caption{Average robustness improvement. \sys-robustified models are 142\% more robust on average compared to Amoeba-robustified ones.}
\label{tab:robustimprov}
\vspace{-0.2cm}
\end{table}

After robustification, we use again \sys, Amoeba and BAP to re-attack (spawn adversarial inputs for) the robustified models $M'$ to re-evaluate models' vulnerability. We train a new agent for Amoeba on $D_{train}'$ to fool each robustified model $M'$. 
Fig.~\ref{fig:val3} shows the 
ASR using \sys, Amoeba, or BAP to find adversarial samples from the vanilla, \sys, and Amoeba robustified models for the end-host threat model (refer to Fig.~\ref{apdx:asr-robust} for the in-path threat model). 
These results show that \sys-robustified models are generally robust against the attack from \sys, Amoeba and BAP, while Amoeba-robustified models are only robust against Amoeba. 
Table~\ref{tab:robustimprov} shows the robustness improvement when robustified by \sys and Amoeba and evaluated with \sys, Amoeba and BAP. 
\sys-robustified models are 52.72\% more robust on average than the vanilla ones while Amoeba-robustified ones are only 21.76\% more robust. 
\sys-robustified models outperform Amoeba-robustified by 142\%.

\new{
We also compare \sys's ability to robustify \netApps with NetShare, which (similar to \sys) generates synthetic data, albeit non-adversarial. 
Fig.~\ref{fig:netshare_robustification} and Fig.~\ref{fig:apdx:netshare_robustification} show APP's accuracy and ASR for end-host threat model implemented with \sys when using the same amount of data generated by \sys and NetShare for robustification. \final{\sys-robustified models have slightly lower accuracy than NetShare-robustified ones. However, the accuracy of \sys-robustified models is still high, meaning that they are accurate enough against common inputs and they are more robust against adversarial perturbation than NetShare-robustified ones.} 
NetShare can hardly improve the models' robustness.}



\begin{finding}
    {\new{\sys can improve the robustness of \netApps even against threat models outside those used during adversarial training.}}
\end{finding}

\begin{figure}[h]
\centering
\begin{subfigure}{.90\linewidth}
    \centering
    \includegraphics[width=.95\linewidth]{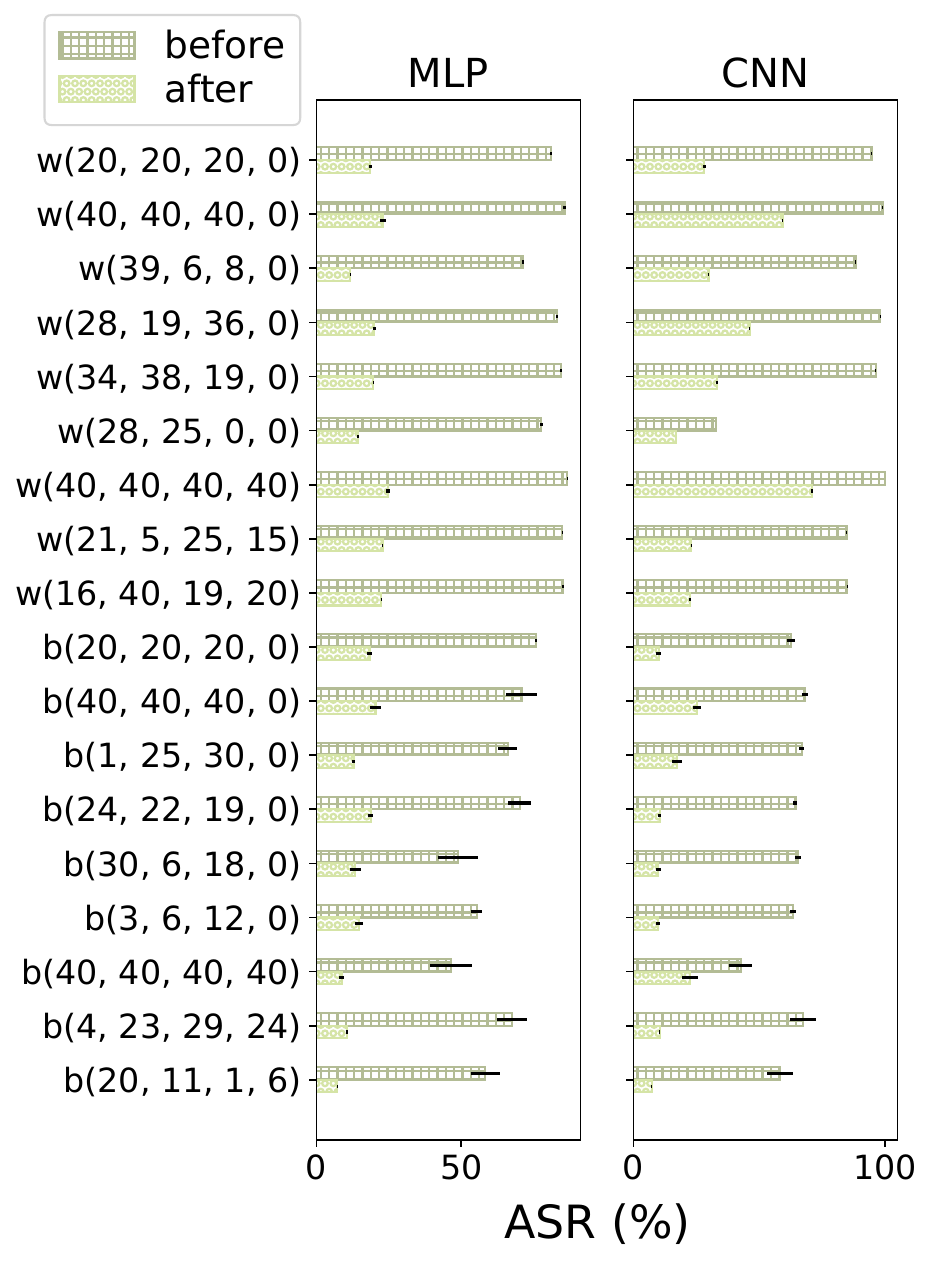}  
    \label{}
\end{subfigure}
\caption{
\new{The ASR of two APP implementations for various threat models, before and after they were robustified by \sys against our end-host attacker (20,20,20,0). \sys improves the robustness of \netApps against threat models that are different or even stronger than what was considered during robustification.
}
}
\label{fig:more-threat}
\end{figure}

\new{
We
evaluate \sys ability to robustify \netApps against a more diverse set of attackers, including attackers that are strictly stronger than what was used during robustification.
To this end, we construct $16$ attackers, shown in Fig.~\ref{fig:more-threat},  which vary on the attacker's access to the model and their capabilities.  The notation w(b) refers to the white-box(black-box) attacker, while the ($\alpha$, $\beta$, $\gamma$, $\delta$) represents an attacker who can delay packets with at most $\alpha\%$, append payload for at most $\beta\%$ packets, inject at most $\gamma$ packets, and split at most $\delta\%$ packets.\footnote{For each packet, we only allow to split at most once to prevent a packet split into too many small fragments.} 
Fig.~\ref{fig:more-threat} reports the ASR for MLP and CNN for the application APP before and after each model has been robustified by \sys. Observe that all models are robustified against the end-host attackers described in~\S\ref{subsec:setup} (\ie a (20, 20, 20, 0) attacker), meaning that many of the attackers tested are stronger.  
Importantly, \sys is able to improve the robustness of these models even against strictly stronger attackers. For instance, \sys-robustified MLP is 72\% more robust than the vanilla version against an attacker that can delay, inject and append twice the amount of the attacker used for robustification and can also split packets.
This happens because during adversarial training, even assuming a weaker or different adversary, the model is guided away from non-robust features (\ie features that are not strongly correlated with the output but rather coincidental or dataset-specific) towards meaningful and resilient features which encapsulate genuine semantics~\cite{ilyas2019adversarial}. 
}


\change{
While \sys-robustified models demonstrate considerable robustness improvement against \sys, Amoeba and BAP (as discussed in Findings 3 and 4), we do not claim that \sys-robustified models are robust against any possible attacks. First, the ASR for \sys-robustified models doesn't drop to 0, meaning that these attack methods can still find adversarial samples against the robustified models, albeit less. Second, \sys assessment is empirical and does not provide theoretical guarantees that the robustified models are robust against the full spectrum of perturbation. 
}


\begin{finding}
    {\new{PANTS is practical and efficient to use, generating \textasciitilde 1.7 samples/sec, which is 8X faster than Amoeba.}}
\end{finding}

\new{
\begin{figure}[h]
\centering
\begin{subfigure}{.89\linewidth}
    \centering
    \includegraphics[width=.9\linewidth]{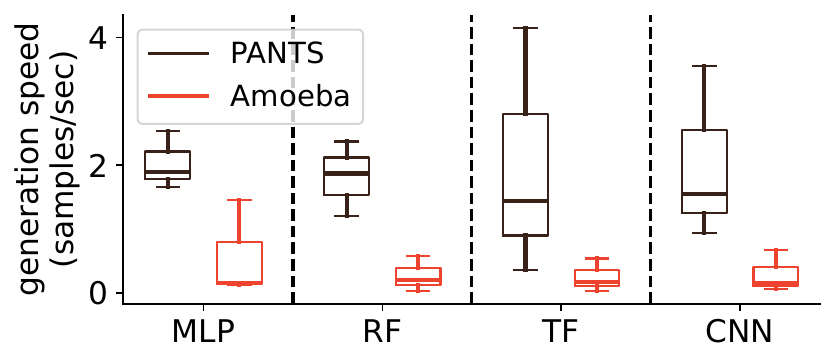} 
    \caption{End-host attacker}
\end{subfigure}
\begin{subfigure}{.89\linewidth}
    \centering
    \includegraphics[width=.9\linewidth]{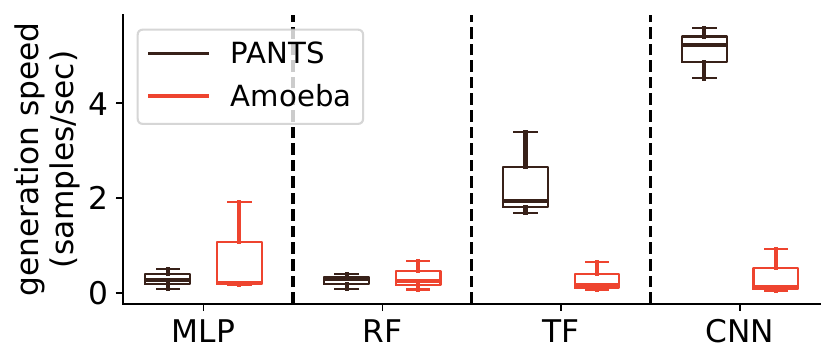} 
    \caption{In-path attacker}
\end{subfigure}
\caption{\new{The generation speed for \sys and Amoeba under different ML models, threat models and applications. PANTS is 8X faster than Ameoba when generating samples.}}
\label{fig:insight2-speed}
\vspace{-0.3cm}
\end{figure}

Thanks to \sys optimizations for improving the scalability of the SMT solver, including chunking, parallelization, and capability approximations as discussed in \S\ref{subsec:smt-opt}, \sys is both practical and efficient.
Fig.~\ref{fig:insight2-speed} shows the generation speed for \sys and Amoeba under our two threat models and all applications and implementations. \sys generates \textasciitilde 1.7 samples/sec, that is  8X faster than Ameoba which generates \textasciitilde 0.2 samples/sec. 
We acknowledge that this comparison is nuanced, as (unlike \sys) Amoeba requires training a model prior to generating adversarial samples. To ensure fairness, we report the amortized time, for generating adversarial examples for as many inputs as there are in each dataset and we run experiments in exactly the same machines (described in Appendix in~\S\ref{apdx:sec:testbed}). 
While the overall speed of \sys clearly outperforms Amoeba, we observe that \sys's speed for MLP and RF under the in-path threat model is relatively slow (\textasciitilde0.3 samples/sec). This is because the in-path threat model, together with complex feature engineering, introduces a more complex set of constraints and narrows the range of potential solutions.
}


\begin{finding}
    {Transferability provides an opportunity to robustify a \netApp without white-box access.}
\end{finding}

\begin{figure}[h]
\centering
\begin{subfigure}{.89\linewidth}
    \centering
    \includegraphics[width=.85\linewidth]{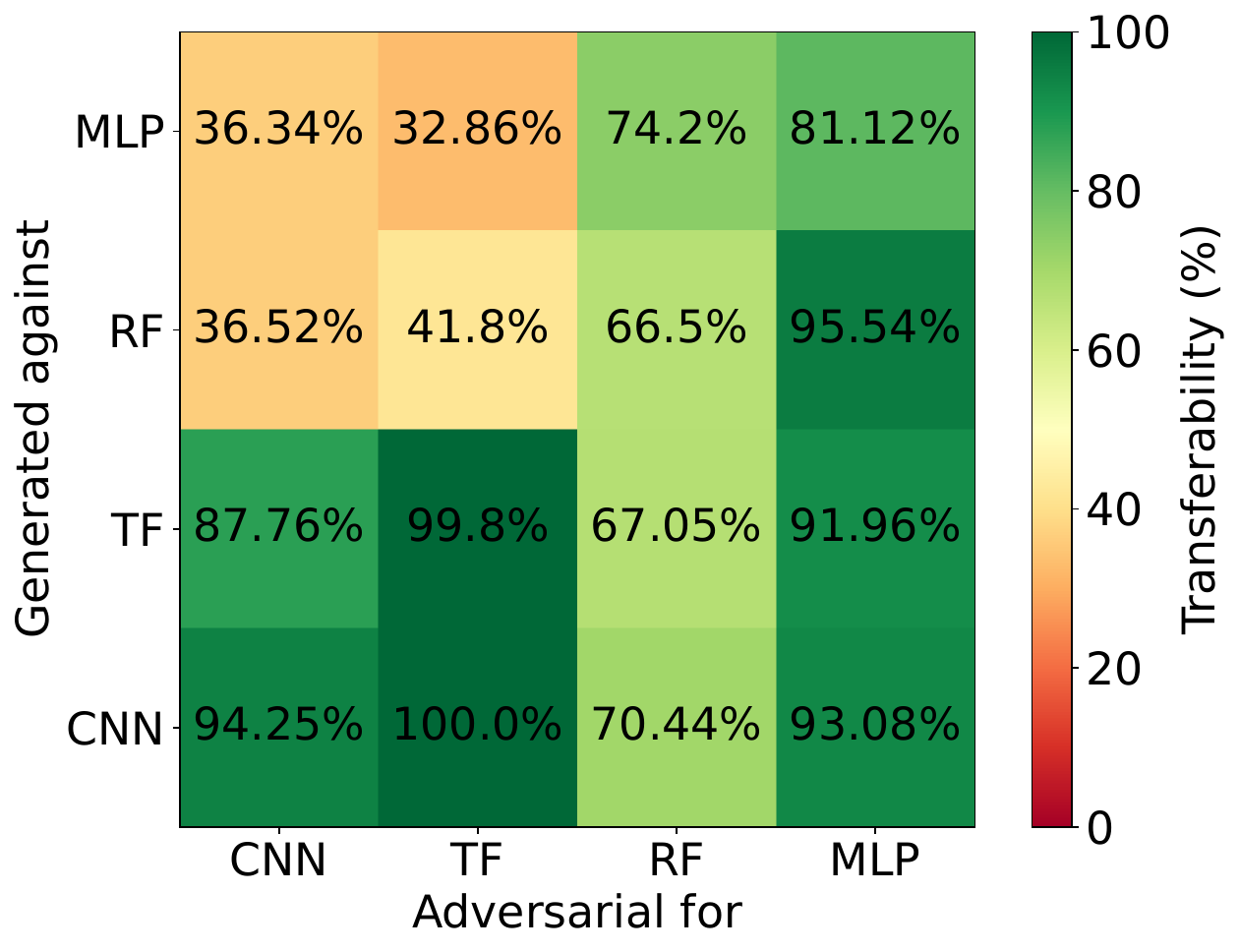}  
\end{subfigure}
\caption{\change{Transferability} results for end-host threat model for the application APP using \sys' generated adversarial samples. Most of the ASR are higher than 50\%, indicating that the adversarial samples generated for one implementation of an \netApp are adversarial for others.}
\label{fig:transfer}
\end{figure}

There are use cases in which a network operator will not have access to the \netApp, \eg because the \netApp is designed by a third party and its details are secret. 
While this is not our target setting, thanks to the transferability of adversarial samples \sys might be able to help the network operator \eg by providing examples in which the target \netApp misclassified. 

\change{Transferability} is the fraction of adversarial samples, crafted for one implementation of a \netApp, that successfully deceive another implementation.
 Concretely, suppose we evaluate the transferability of the adversarial samples generated from $M_A$ to $M_B$, we first construct the test set $X_{test\_A\_B}'\subseteq X_{test}$ that both $M_A$ and $M_B$ can correctly classify. Then we use \sys to generate adversarial samples for $X_{test\_A\_B}'$ using $M_A$ and evaluate its adversariness against $M_B$. Fig.~\ref{fig:transfer} shows the transferability result between MLP, RF, TF and CNN for end-host threat model tested for the application APP.  Most of the ASR (outside the diagonal) is above 50\%, showing that \sys can be useful even without full access to the \netApp implementation.
To benefit from it, the operator would need to train a proxy model and find adversarial samples against it using \sys. According to Fig.~\ref{fig:transfer}, the adversarial samples against the proxy model are likely to be adversarial against the target one. 


%% file: sections/8-conclusion.tex
\vspace{-0.3cm}
\section{Conclusion}
\vspace{-0.2cm}
This paper presents \sys, a practical framework to help network operators evaluate and enhance the robustness of ML-powered networking classifiers. \sys combines canonical AML methods, such as PGD, with an SMT solver to generate adversarial, realizable, and semantic-preserving packet sequences with a high success rate.
%
Our comprehensive evaluation in three \netApps across multiple implementations shows that \sys outperforms baselines in generating adversarial samples by 70\% and 2X in ASR. Importantly, by integrating \sys in an iterative adversarial training pipeline, we find that \sys can robustify \netApps against various attack methodologies by 52.72\% on average.

%% file: sections/acknowledgement.tex
\final{
\section*{Acknowledgement}

We thank the anonymous reviewers for their informative and insightful feedback.
This work was supported by the National Science Foundation (NSF) through Grant CNS-2319442, and a Princeton NextG Innovation Award.
}

%% file: sections/ethics.tex
\new{
\section*{Ethics Considerations}

We evaluate PANTS and all baselines on public datasets avoiding any privacy concerns. PANTS and other baselines are not used to perturb live network traffic or an online system, thus avoiding ethical concerns. Our primary goal in proposing PANTS is to assist network operators in evaluating and enhancing the robustness of MNCs to defend against adversaries.
}

%% file: sections/open-science.tex
\new{
\section*{Open Science}
\label{sec:open-science}

We release all artifacts.

\final{First, we release all code and scripts from data processing, MNC training and inference, and PANTS (SMT, AML, including integration with PGD and ZOO)  via both Zenodo (\url{https://zenodo.org/records/14728507}) and GitHub (\url{https://github.com/jinminhao/PANTS}).} In terms of secondary artifacts, we share all parameters and configurations used for MNC training and PANTS. We also release the trained MNC models, including both vanilla and robustified versions.

Furthermore, we provide adversarial samples generated by PANTS for each application and model as references. Any artifacts related to the model and data are shared via Zenodo.

We already use public datasets to ensure reproducibility, and the references to these datasets are provided in the paper. That said, we also release the dataset after pre-preprocessing.

}

%% file: sections/appendix.tex
\appendix
\section{Robustification via synthetic data augmentation}
\label{apdx:sec:syn}
\begin{figure}[h]
\centering
\begin{subfigure}{.475\linewidth}
    \centering
    \includegraphics[width=.95\linewidth]{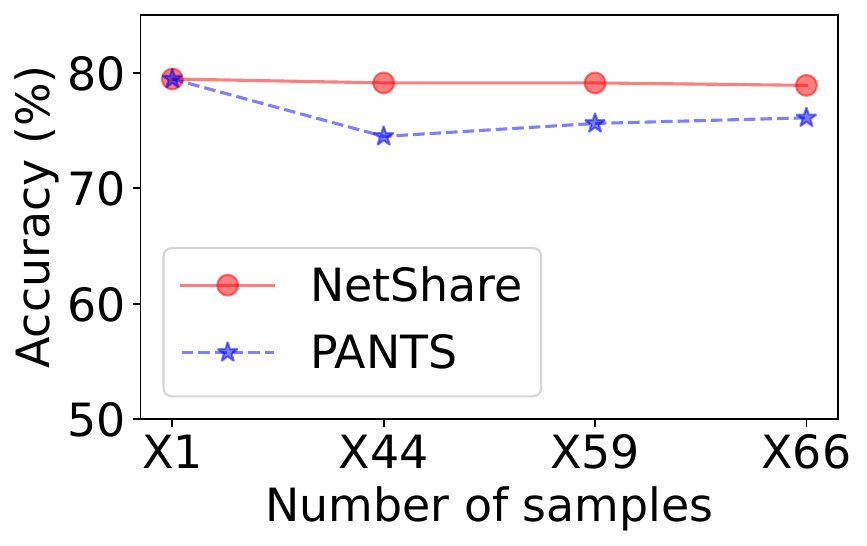}  
    \caption{Accuracy of MLP}
    \label{}
\end{subfigure}
\begin{subfigure}{.475\linewidth}
    \centering
    \includegraphics[width=.95\linewidth]{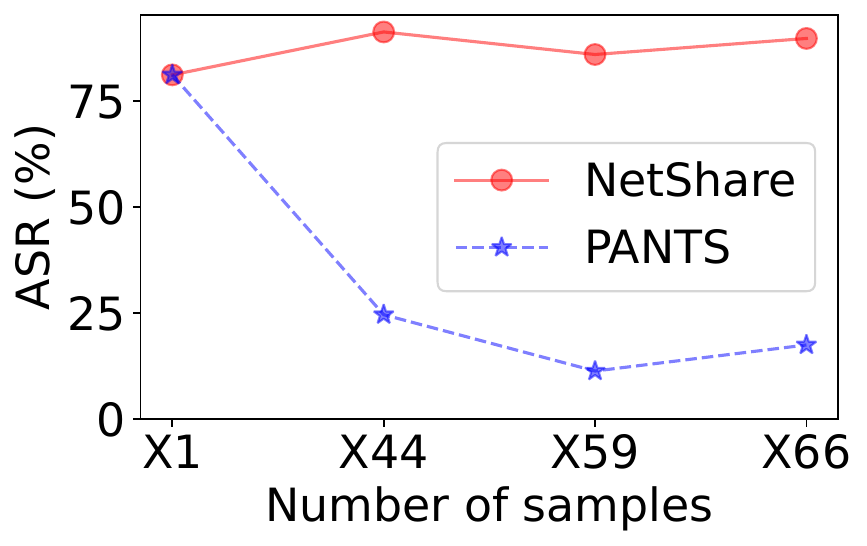}  
    \caption{Vulnerability of MLP}
    \label{}
\end{subfigure}
\begin{subfigure}{.475\linewidth}
    \centering
    \includegraphics[width=.95\linewidth]{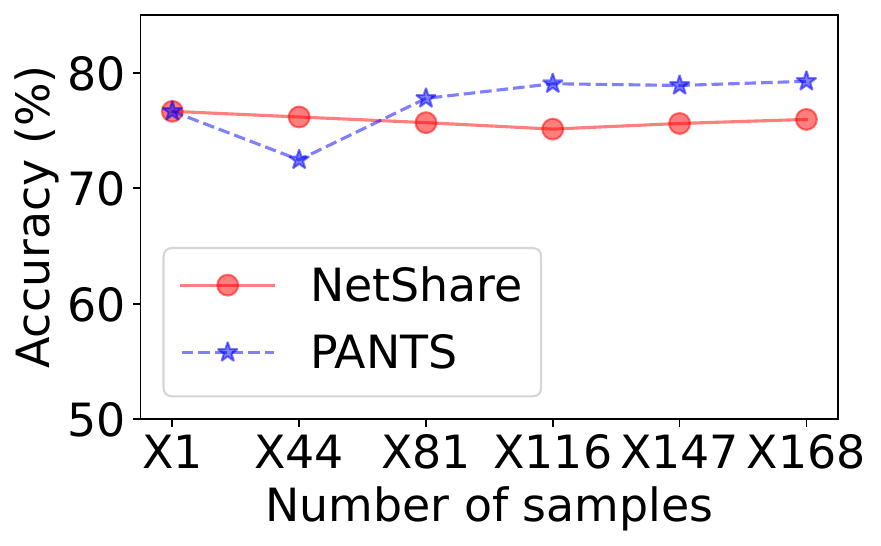}  
    \caption{Accuracy of TF}
    \label{}
\end{subfigure}
\begin{subfigure}{.475\linewidth}
    \centering
    \includegraphics[width=.95\linewidth]{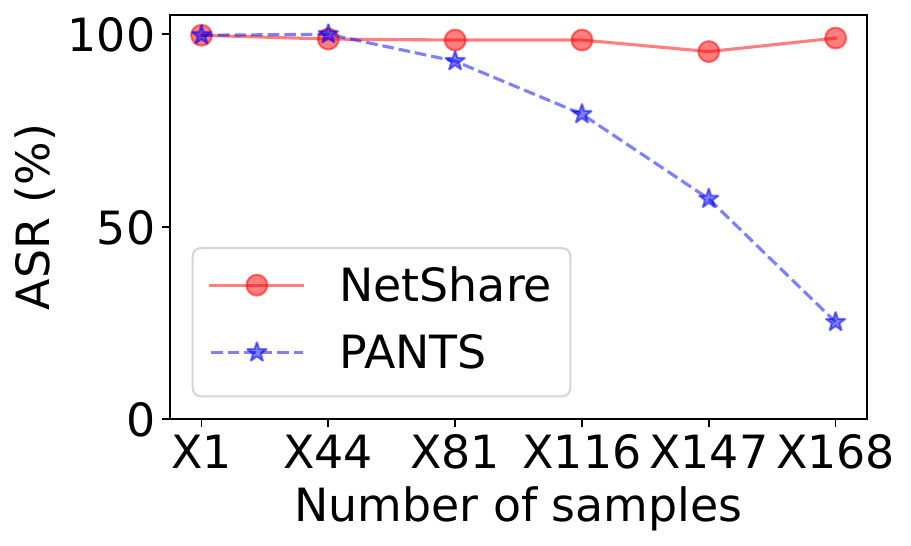}  
    \caption{Vulnerability of TF}
    \label{}
\end{subfigure}
\caption{\new{
Model (MLP and TF) accuracy and ASR when using synthetic data by NetShare and adversarial samples by \sys for robustification.
}
}
\label{fig:apdx:netshare_robustification}
\end{figure}

\new{Fig.~\ref{fig:apdx:netshare_robustification} shows the accuracy and vulnerability of MLP and transformer when robustifying using synthetic data generated by NetShare.}

\section{Features used in different applications}
\label{apdx:sec:feat}
Different features are extracted from the feature engineering module among applications. Table~\ref{apdx:tab:dataset} presents a summary of the
features employed for each dataset. The selection of features
is following the corresponding paper or the given script from
the released dataset.
\begin{table*}[h]
\centering
\begin{tabular}{l|l}
\hline
Application                                                          & Features                                                                                                                                                                                                                                       \\ \hline
VPN Detection  (VPN)                                                      & \begin{tabular}[c]{@{}l@{}}\{sum, min, max, mean\} of \{bi-directional, forward, backward\} flow interarrival times\\ \{min, max, mean, std\} of \{active, idle\} durations\\ Number of \{packets, bytes\} per second\end{tabular}                             \\ \hline
\begin{tabular}[c]{@{}l@{}}Application\\ Identification (APP)\end{tabular} & \begin{tabular}[c]{@{}l@{}}\{sum, min, max, mean, std\} of \{forward, backward\} flow interarrival times\\ \{sum, min, max, mean, std\} of packet length of the \{forward, backward\} flow\\ Number of total \{forward, backward\} packets\end{tabular} \\ \hline
\begin{tabular}[c]{@{}l@{}}VCAs QoE\\ Inference (QOE)\end{tabular}        & \begin{tabular}[c]{@{}l@{}}\{min, max, mean, median, std\} of \{packet length, interarrival times\}\\ Number of \{bytes, packets\}\\ Number of \{unique packet length, microbursts\}\end{tabular}  \\     \hline                                     
\end{tabular}
\caption{Description of the used dataset and features extracted by the feature engineering module for each application. An active duration is defined as the amount of time that flow is active before going to idle. An idle duration is defined vice versa\cite{iscxvpn2016}. A flow is active when the interarrival times are less than a threshold (0.1s in our case). }
\label{apdx:tab:dataset}
\end{table*}

\section{Hyperparemeters of ML models}
\label{apdx:sec:hp}
To simplify the application setup, for each kind of ML classifier, we use the same architecture and hyperparameters for all the applications.  The MLP is set with 4 hidden layers with 200 neurons each. It is trained with 500 epochs, and the optimizer is SGD. Since VPN detection is binary classification, a sigmoid function is used as the last activation layer, and the loss function is defined as binary cross entropy. Other applications are multi-class classification, hence we are using softmax function and the loss function is cross entropy. RF is implemented by Scikit-learn~\cite{sklearn}. The maximum depth of trees is set as 12 and other parameters are set as default. The TF is borrowed from the idea of BERT~\cite{devlin2018bert}. It has 4 attention heads, 2 hidden layers and the intermediate size is 256. The CNN is derived from DF~\cite{sirinam2018deep} which is previously designed for website fingerprinting. It has four sequential blocks where each includes two convolution layers and one max pooling layer, followed by three fully-connected layers. For both TF and CNN, only the first 400 packets for each packet sequence are encoded into the input. Packet sequences whose length are less than 400 will be padded with 0 for input format consistency. Both of them are trained with 200 epochs.

\section{Capability extension of Amoeba}
\label{apdx:sec:amoebaext}
For a fair comparison, we extend Amoeba's capability to comply with the two evaluated threat models.  First, we identify perturbable packets. Since the attacker's capability can only perturb a single direction of packets instead of bi-direction, we modify the action function to work on either forward or backward packets.  Then, we extend the action space of Amoeba to support more perturbations. The original action space has two features, representing the delay and the modified packet length of the given packet. We extend it by adding an integer to represent the number of injected packets after this packet. When the threat model assumes an end-host attacker, this newly added feature executes after each packet. We find that most of the adversarial samples generated by Amoeba are in conflict with the semantics constraints such as constraint of accumulative delay and maximum number of injected packets. We just add a hard budget to force the agent not to break this constraints. We found that this would help to improve Amoeba's ASR compared to simply dropping all the adversarial sequences which break these flow-wise constraints. Further, for the applications that require multi-class classification,  we modify the reward function. Amoeba's original reward function only considers binary classification. Therefore, when targeting multi-class classification, we changed the reward function to cross-entropy. The agent can get a large reward when perturbing the original sample to cause misclassification.

\section{Testbed specification}
\label{apdx:sec:testbed}
For consistent evaluation, all the experiments using \sys to generate adversarial samples
are running on the cluster of Cloudlab machines~\cite{cloudlab}. Each machine has two Intel Xeon Silver 4114 10-core CPUs at 2.20 GHz and 192GB ECC DDR4-2666 memory. The rest of the experiments related to Amoeba rely on GPU training, which is running on two Azure machines~\cite{azure_cloud}. Each one uses 16 vCPUs from AMD EPYC 7V12(Rome) CPU, 110GB memory, and an Nvidia Tesla T4 GPU with 16 GB GPU memory.  \change{Please note that the training and inference of Amoeba is on the CPU machines (same as \sys) when comparing the computation overhead with \sys.}
\section{Robustness of the \sys- and Amoeba-robustified models}
\begin{figure*}[h]
\centering
\begin{subfigure}{.8\textwidth}
    \centering
    \includegraphics[width=.85\linewidth]{figures/eval_3_legend.pdf}  
\end{subfigure}
\begin{subfigure}{.33\textwidth}
    \centering
    \includegraphics[width=.95\linewidth]{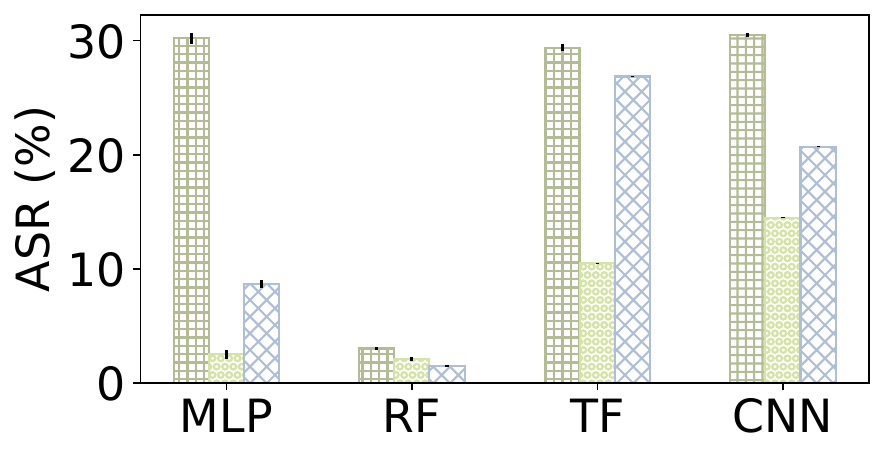}  
    \caption{APP Attacked by \sys}
    \label{}
\end{subfigure}
\begin{subfigure}{.33\textwidth}
    \centering
    \includegraphics[width=.95\linewidth]{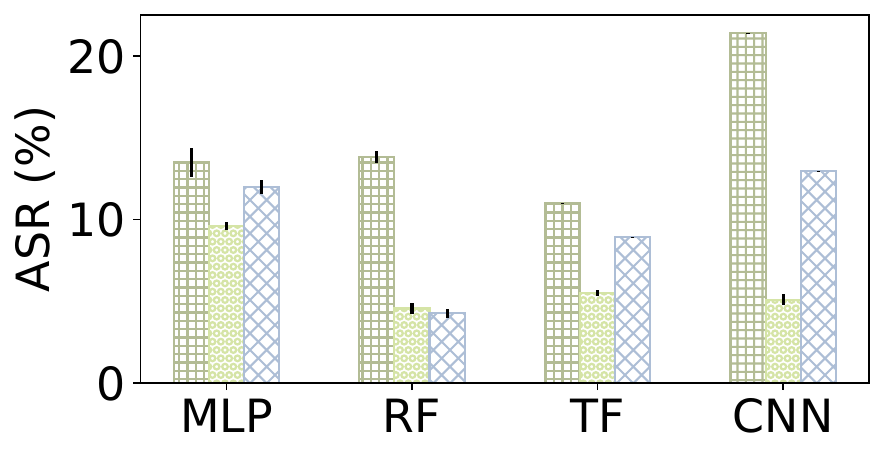}  
    \caption{VPN Attacked by \sys}
    \label{}
\end{subfigure}
\begin{subfigure}{.33\textwidth}
    \centering
    \includegraphics[width=.95\linewidth]{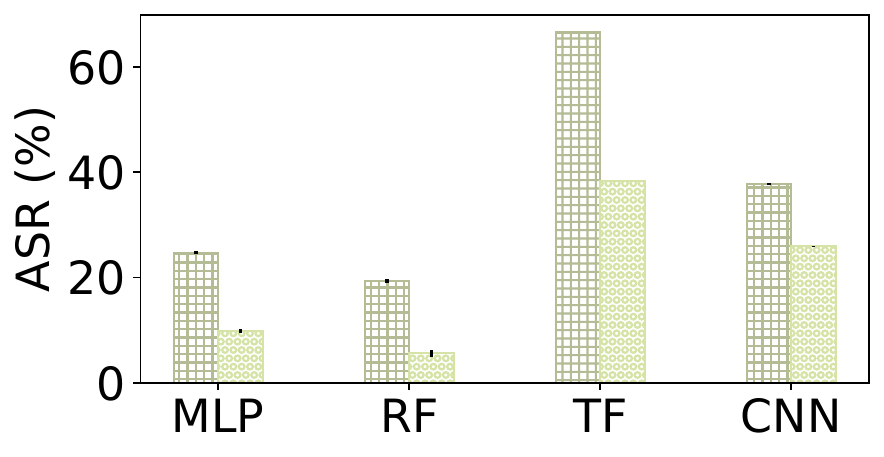}  
    \caption{QOE Attacked by \sys}
    \label{}
\end{subfigure}
\begin{subfigure}{.33\textwidth}
    \centering
    \includegraphics[width=.95\linewidth]{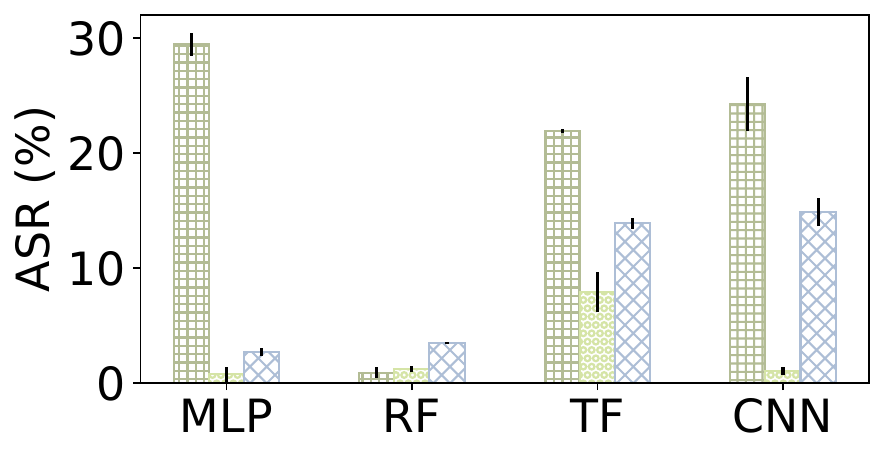}  
    \caption{APP Attacked by Amoeba}
    \label{}
\end{subfigure}
\begin{subfigure}{.33\textwidth}
    \centering
    \includegraphics[width=.95\linewidth]{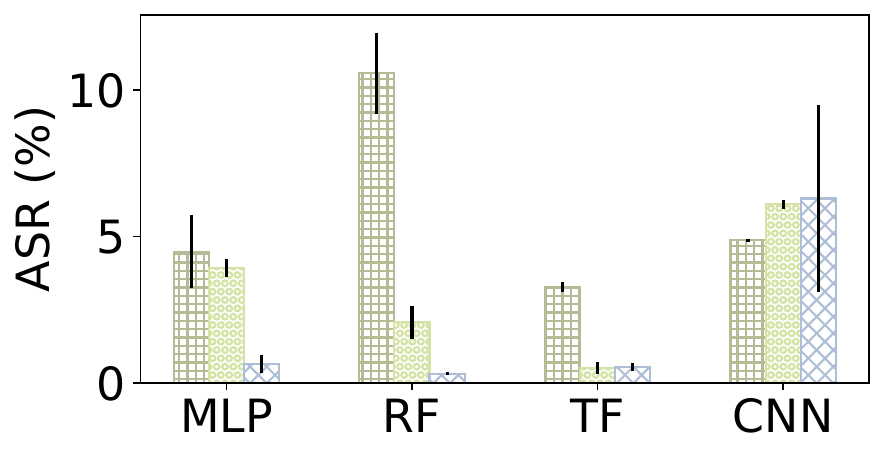}  
    \caption{VPN Attacked by Amoeba}
    \label{}
\end{subfigure}
\begin{subfigure}{.33\textwidth}
    \centering
    \includegraphics[width=.95\linewidth]{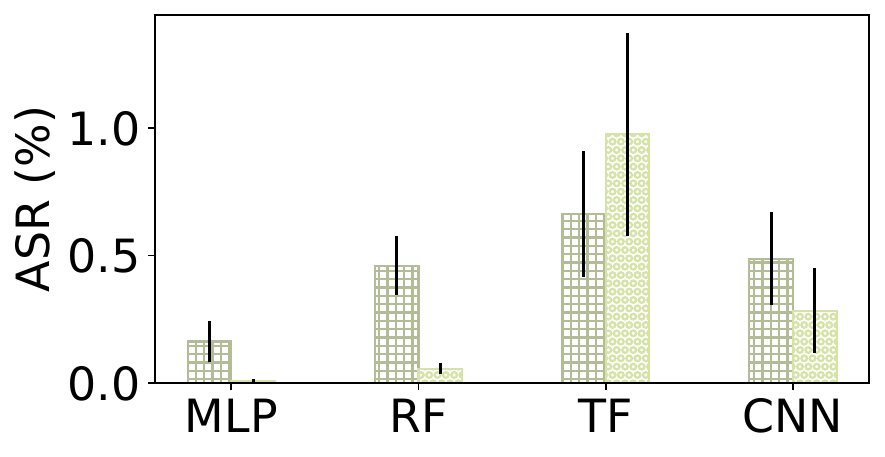}  
    \caption{QOE Attacked by Amoeba}
    \label{apdx:asr-robust-vca-weak-amoeba}
\end{subfigure}
\begin{subfigure}{.22\textwidth}
    \centering
    \includegraphics[width=.95\linewidth]{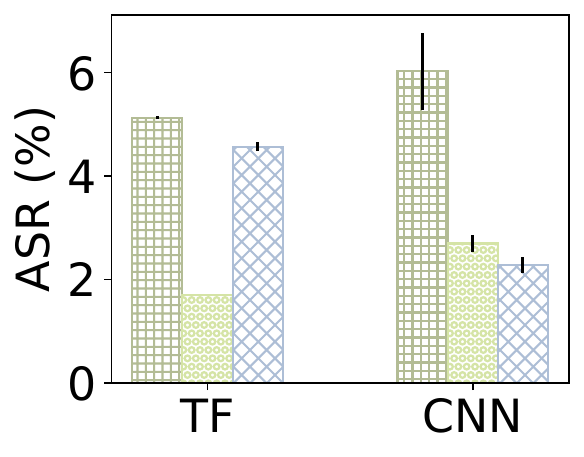}  
    \caption{APP Attacked by BAP}
    \label{}
\end{subfigure}
\begin{subfigure}{.22\textwidth}
    \centering
    \includegraphics[width=.95\linewidth]{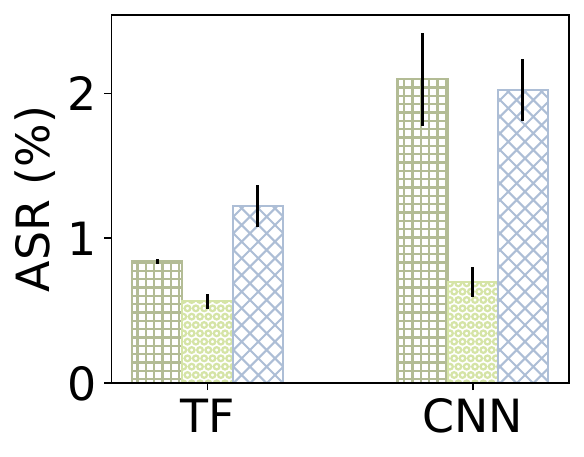}
    \caption{VPN Attacked by BAP}
    \label{}
\end{subfigure}
\begin{subfigure}{.22\textwidth}
    \centering
    \includegraphics[width=.95\linewidth]{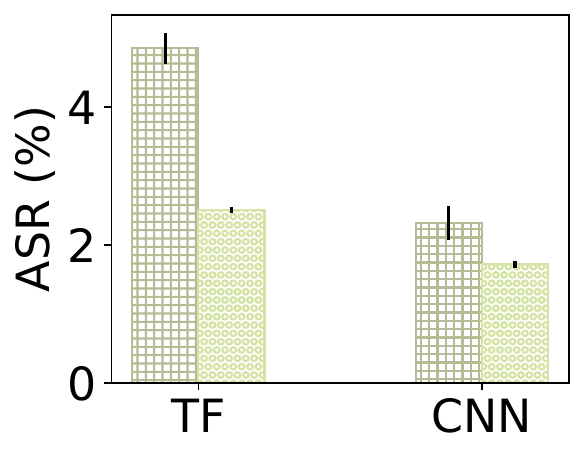}  
    \caption{QOE Attacked by BAP}
    \label{}
\end{subfigure}
\caption{ASR for the vanilla, PANTS-robustified and Amoeba-robustified models for the \change{\textbf{in-path}} threat model. \new{Similar to the observations in fig.~\ref{fig:val3},} PANTS-robustified’s ASR is lower, demonstrating PANTS’s ability to generate adversarial samples that effectively robustify target models. For the dataset QOE, using Amoeba for robustification is skipped since it cannot effectively generate enough adversarial samples.}
\label{apdx:asr-robust}
\end{figure*}

Fig.~\ref{apdx:asr-robust} shows the robustness of using \sys, Amoeba and BAP to attack \sys- and Amoeba-robustified models under the in-path threat model.

\begin{figure}[h]
\centering
\begin{subfigure}{.475\linewidth}
    \centering
    \includegraphics[width=.95\linewidth]{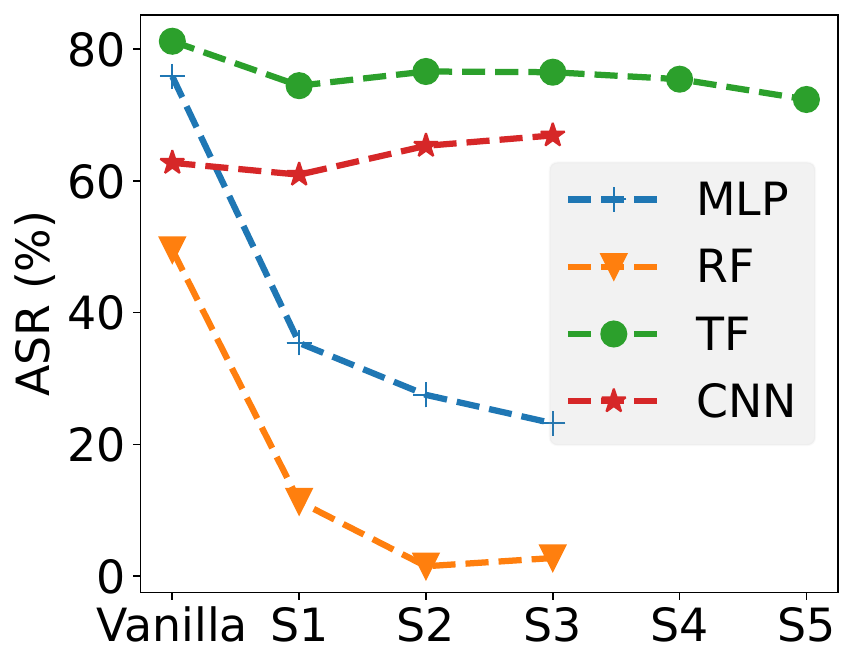}  
    \caption{Robustified for defending end-host attackers}
    \label{app-strong}
\end{subfigure}
\begin{subfigure}{.475\linewidth}
    \centering
    \includegraphics[width=.95\linewidth]{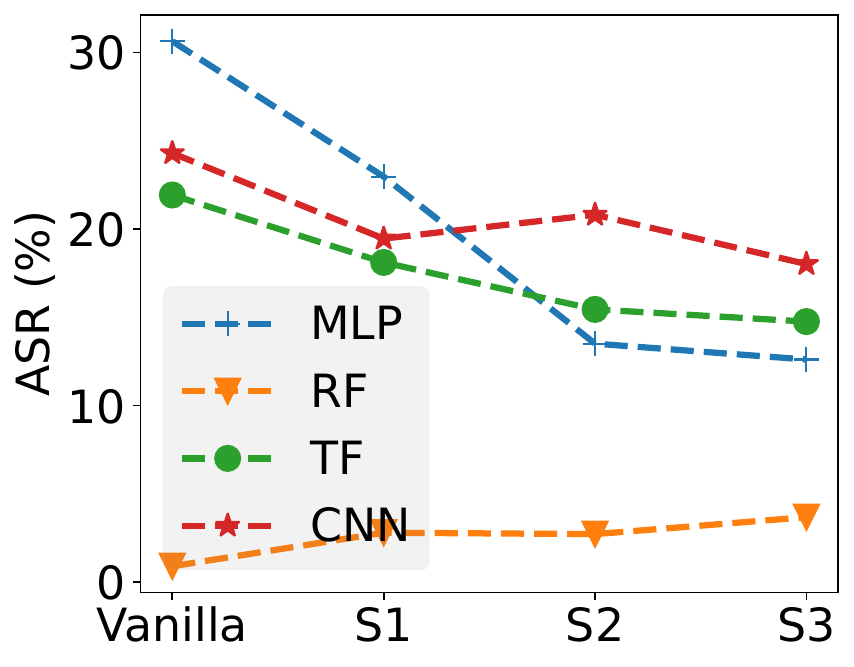}  
    \caption{Robustified for defending in-path attackers}
    \label{app-strong}
\end{subfigure}
\caption{ASR by using Amoeba to attack the ML model
when it is in iterative Amoeba-robustification process.}
\label{fig:robust_amoeba_asr_drop}
\end{figure}

%% file: main.bbl
\begin{thebibliography}{10}

\bibitem{artszoorf}
Adversarial-robustness-toolbox for scikit-learn randomforestclassifier.
\newblock \url{https://github.com/Trusted-AI/adversarial-robustness-toolbox/blob/main/notebooks/classifier_scikitlearn_RandomForestClassifier.ipynb}.

\bibitem{artszooxgb}
Adversarial-robustness-toolbox for xgboost.
\newblock \url{https://github.com/Trusted-AI/adversarial-robustness-toolbox/blob/main/notebooks/classifier_xgboost.ipynb}.

\bibitem{azure_cloud}
azure.
\newblock \url{https://azure.microsoft.com/}.

\bibitem{CAIDA}
The caida ucsd anonymized internet traces.
\newblock \url{https://www.caida.org/catalog/datasets/passive_dataset}.
\newblock Accessed: 2022-01-30.

\bibitem{att}
Harnessing data and ai for business value.
\newblock \url{https://about.att.com/innovationblog/2022/data-ai-part-1.html}.

\bibitem{juniper}
Mist ai and cloud.
\newblock \url{https://www.juniper.net/us/en/products/mist-ai.html}.

\bibitem{sklearn}
scikit-learn.
\newblock \url{https://scikit-learn.org/stable/}.

\bibitem{cisco}
Solution overview artificial intelligence/machine learning for intent-based networking – primer.
\newblock \url{https://www.cisco.com/c/en/us/solutions/collateral/enterprise-networks/digital-network-architecture/nb-06-cisco-dna-ai-ml-primer-cte-en.html}.

\bibitem{Alahmed2022zoorf}
Shahad Alahmed, Qutaiba Alasad, Maytham~M. Hammood, Jiann-Shiun Yuan, and Mohammed Alawad.
\newblock Mitigation of black-box attacks on intrusion detection systems-based ml.
\newblock {\em Computers}, 11(7), 2022.

\bibitem{Aneja2023iot}
Nagender Aneja, Sandhya Aneja, and Bharat Bhargava.
\newblock Ai-enabled learning architecture using network traffic traces over iot network: A comprehensive review.
\newblock {\em Wireless Communications and Mobile Computing}, 2023:8658278, Feb 2023.

\bibitem{benson2010network}
Theophilus Benson, Aditya Akella, and David~A Maltz.
\newblock Network traffic characteristics of data centers in the wild.
\newblock In {\em Proceedings of the 10th ACM SIGCOMM conference on Internet measurement}, pages 267--280, 2010.

\bibitem{carlini2017towards}
Nicholas Carlini and David Wagner.
\newblock Towards evaluating the robustness of neural networks.
\newblock In {\em 2017 ieee symposium on security and privacy (sp)}, pages 39--57. Ieee, 2017.

\bibitem{Carofiglio2021Cha}
Giovanna Carofiglio, Giulio Grassi, Enrico Loparco, Luca Muscariello, Michele Papalini, and Jacques Samain.
\newblock Characterizing the relationship between application qoe and network qos for real-time services.
\newblock In {\em Proceedings of the ACM SIGCOMM 2021 Workshop on Network-Application Integration}, NAI'21, page 20–25, New York, NY, USA, 2021. Association for Computing Machinery.

\bibitem{chen2017zoo}
Pin-Yu Chen, Huan Zhang, Yash Sharma, Jinfeng Yi, and Cho-Jui Hsieh.
\newblock Zoo: Zeroth order optimization based black-box attacks to deep neural networks without training substitute models.
\newblock In {\em Proceedings of the 10th ACM workshop on artificial intelligence and security}, pages 15--26, 2017.

\bibitem{devlin2018bert}
Jacob Devlin.
\newblock Bert: Pre-training of deep bidirectional transformers for language understanding.
\newblock {\em arXiv preprint arXiv:1810.04805}, 2018.

\bibitem{10.5555/550359}
Edsger~Wybe Dijkstra.
\newblock {\em A Discipline of Programming}.
\newblock Prentice Hall PTR, USA, 1st edition, 1997.

\bibitem{Doshi2018machine}
Rohan Doshi, Noah Apthorpe, and Nick Feamster.
\newblock Machine learning ddos detection for consumer internet of things devices.
\newblock In {\em 2018 IEEE Security and Privacy Workshops (SPW)}, pages 29--35, 2018.

\bibitem{iscxvpn2016}
Gerard Draper-Gil, Arash~Habibi Lashkari, Mohammad Saiful~Islam Mamun, and Ali A.~Ghorbani.
\newblock Characterization of encrypted and vpn traffic using time-related features.
\newblock In {\em Proceedings of the 2nd International Conference on Information Systems Security and Privacy}, pages 407--414, 2016.
\newblock https://doi.org/10.5220/0005740704070414.

\bibitem{cloudlab}
Dmitry Duplyakin, Robert Ricci, Aleksander Maricq, Gary Wong, Jonathon Duerig, Eric Eide, Leigh Stoller, Mike Hibler, David Johnson, Kirk Webb, Aditya Akella, Kuangching Wang, Glenn Ricart, Larry Landweber, Chip Elliott, Michael Zink, Emmanuel Cecchet, Snigdhaswin Kar, and Prabodh Mishra.
\newblock The design and operation of {CloudLab}.
\newblock In {\em 2019 USENIX Annual Technical Conference (USENIX ATC 19)}, pages 1--14, Renton, WA, July 2019. USENIX Association.

\bibitem{floyd1993assigning}
Robert~W Floyd.
\newblock Assigning meanings to programs.
\newblock In {\em Program Verification: Fundamental Issues in Computer Science}, pages 65--81. Springer, 1993.

\bibitem{VPN2020sequence}
Ping Gao, Guangsong Li, Yanan Shi, and Yang Wang.
\newblock Vpn traffic classification based on payload length sequence.
\newblock In {\em 2020 International Conference on Networking and Network Applications (NaNA)}, pages 241--247, 2020.

\bibitem{tomer2019robustify}
Tomer Gilad, Nathan~H. Jay, Michael Shnaiderman, Brighten Godfrey, and Michael Schapira.
\newblock Robustifying network protocols with adversarial examples.
\newblock In {\em Proceedings of the 18th ACM Workshop on Hot Topics in Networks}, HotNets '19, page 85–92, New York, NY, USA, 2019. Association for Computing Machinery.

\bibitem{goodfellow2014explaining}
Ian~J Goodfellow, Jonathon Shlens, and Christian Szegedy.
\newblock Explaining and harnessing adversarial examples.
\newblock {\em arXiv preprint arXiv:1412.6572}, 2014.

\bibitem{goodfellow2014fgsm}
Ian~J Goodfellow, Jonathon Shlens, and Christian Szegedy.
\newblock Explaining and harnessing adversarial examples.
\newblock {\em arXiv preprint arXiv:1412.6572}, 2014.

\bibitem{grabinski2022robust}
Julia Grabinski, Paul Gavrikov, Janis Keuper, and Margret Keuper.
\newblock Robust models are less over-confident.
\newblock {\em Advances in Neural Information Processing Systems}, 35:39059--39075, 2022.

\bibitem{utmobile}
Yuqiang Heng, Vikram Chandrasekhar, and Jeffrey~G. Andrews.
\newblock Utmobilenettraffic2021: A labeled public network traffic dataset.
\newblock {\em IEEE Networking Letters}, 3(3):156--160, 2021.

\bibitem{Hossain2017quality}
Md. Faisal~Murad Hossain, Mahasweta Sarkar, and Syed~Hassan Ahmed.
\newblock Quality of experience for video streaming: A contemporary survey.
\newblock In {\em 2017 13th International Wireless Communications and Mobile Computing Conference (IWCMC)}, pages 80--84, 2017.

\bibitem{ilyas2019adversarial}
Andrew Ilyas, Shibani Santurkar, Dimitris Tsipras, Logan Engstrom, Brandon Tran, and Aleksander Madry.
\newblock Adversarial examples are not bugs, they are features.
\newblock {\em Advances in neural information processing systems}, 32, 2019.

\bibitem{jiang2016via}
Junchen Jiang, Rajdeep Das, Ganesh Ananthanarayanan, Philip~A Chou, Venkata Padmanabhan, Vyas Sekar, Esbjorn Dominique, Marcin Goliszewski, Dalibor Kukoleca, Renat Vafin, et~al.
\newblock Via: Improving internet telephony call quality using predictive relay selection.
\newblock In {\em Proceedings of the 2016 ACM SIGCOMM Conference}, pages 286--299. ACM, 2016.

\bibitem{jiang2016cfa}
Junchen Jiang, Vyas Sekar, Henry Milner, Davis Shepherd, Ion Stoica, and Hui Zhang.
\newblock Cfa: A practical prediction system for video qoe optimization.
\newblock In {\em NSDI}, pages 137--150, 2016.

\bibitem{netdifJiang}
Xi~Jiang, Shinan Liu, Aaron Gember-Jacobson, Arjun~Nitin Bhagoji, Paul Schmitt, Francesco Bronzino, and Nick Feamster.
\newblock Netdiffusion: Network data augmentation through protocol-constrained traffic generation.
\newblock {\em Proc. ACM Meas. Anal. Comput. Syst.}, 8(1), feb 2024.

\bibitem{kumar2022iot}
Rakesh Kumar, Mayank Swarnkar, Gaurav Singal, and Neeraj Kumar.
\newblock Iot network traffic classification using machine learning algorithms: An experimental analysis.
\newblock {\em IEEE Internet of Things Journal}, 9(2):989--1008, 2022.

\bibitem{li2023sokcertified}
Linyi Li, Tao Xie, and Bo~Li.
\newblock Sok: Certified robustness for deep neural networks.
\newblock In {\em 2023 IEEE symposium on security and privacy (SP)}, pages 1289--1310. IEEE, 2023.

\bibitem{lin2022bert}
Xinjie Lin, Gang Xiong, Gaopeng Gou, Zhen Li, Junzheng Shi, and Jing Yu.
\newblock Et-bert: A contextualized datagram representation with pre-training transformers for encrypted traffic classification.
\newblock In {\em Proceedings of the ACM Web Conference 2022}, pages 633--642, 2022.

\bibitem{amoeba}
Haoyu Liu, Alec~F. Diallo, and Paul Patras.
\newblock Amoeba: Circumventing ml-supported network censorship via adversarial reinforcement learning.
\newblock In {\em Proceedings of the 13th International Conference on emerging Networking EXperiments and Technologies (CoNEXT)}, 2023.

\bibitem{TC2017lopez}
Manuel Lopez-Martin, Belen Carro, Antonio Sanchez-Esguevillas, and Jaime Lloret.
\newblock Network traffic classifier with convolutional and recurrent neural networks for internet of things.
\newblock {\em IEEE Access}, 5:18042--18050, 2017.

\bibitem{madry2017pgd}
Aleksander Madry, Aleksandar Makelov, Ludwig Schmidt, Dimitris Tsipras, and Adrian Vladu.
\newblock Towards deep learning models resistant to adversarial attacks.
\newblock {\em arXiv preprint arXiv:1706.06083}, 2017.

\bibitem{VPN2022TPIPD}
Yongwei Meng, Tao Qin, Haonian Wang, and Zhouguo Chen.
\newblock Tpipd: A robust model for online vpn traffic classification.
\newblock In {\em 2022 IEEE International Conference on Trust, Security and Privacy in Computing and Communications (TrustCom)}, pages 105--110, 2022.

\bibitem{TC2005}
Andrew~W. Moore and Denis Zuev.
\newblock Internet traffic classification using bayesian analysis techniques.
\newblock {\em SIGMETRICS Perform. Eval. Rev.}, 33(1):50–60, June 2005.

\bibitem{nandy2021towards}
Jay Nandy, Sudipan Saha, Wynne Hsu, Mong~Li Lee, and Xiao~Xiang Zhu.
\newblock Towards bridging the gap between empirical and certified robustness against adversarial examples.
\newblock {\em arXiv preprint arXiv:2102.05096}, 2021.

\bibitem{BAP}
Milad Nasr, Alireza Bahramali, and Amir Houmansadr.
\newblock Defeating {DNN-Based} traffic analysis systems in {Real-Time} with blind adversarial perturbations.
\newblock In {\em 30th USENIX Security Symposium (USENIX Security 21)}, pages 2705--2722. USENIX Association, August 2021.

\bibitem{Nikravesh2016QoE}
Ashkan Nikravesh, David~Ke Hong, Qi~Alfred Chen, Harsha~V. Madhyastha, and Z.~Morley Mao.
\newblock Qoe inference without application control.
\newblock In {\em Proceedings of the 2016 Workshop on QoE-Based Analysis and Management of Data Communication Networks}, Internet-QoE '16, page 19–24, New York, NY, USA, 2016. Association for Computing Machinery.

\bibitem{parchekani2020classification}
Ali Parchekani, Salar Nouri, Vahid Shah-Mansouri, and Seyed~Pooya Shariatpanahi.
\newblock Classification of traffic using neural networks by rejecting: a novel approach in classifying vpn traffic.
\newblock {\em arXiv preprint arXiv:2001.03665}, 2020.

\bibitem{peng2019mask}
Xiao Peng, Weiqing Huang, and Zhixin Shi.
\newblock Adversarial attack against dos intrusion detection: An improved boundary-based method.
\newblock In {\em 2019 IEEE 31st International Conference on Tools with Artificial Intelligence (ICTAI)}, pages 1288--1295, 2019.

\bibitem{azure}
Olga Poppe, Tayo Amuneke, Dalitso Banda, Aritra De, Ari Green, Manon Knoertzer, Ehi Nosakhare, Karthik Rajendran, Deepak Shankargouda, Meina Wang, et~al.
\newblock Seagull: An infrastructure for load prediction and optimized resource allocation.
\newblock {\em arXiv preprint arXiv:2009.12922}, 2020.

\bibitem{salman2020adversarially}
Hadi Salman, Andrew Ilyas, Logan Engstrom, Ashish Kapoor, and Aleksander Madry.
\newblock Do adversarially robust imagenet models transfer better?
\newblock {\em Advances in Neural Information Processing Systems}, 33:3533--3545, 2020.

\bibitem{sharma2022lumen}
Rahul~Anand Sharma, Ishan Sabane, Maria Apostolaki, Anthony Rowe, and Vyas Sekar.
\newblock Lumen: a framework for developing and evaluating ml-based iot network anomaly detection.
\newblock In {\em Proceedings of the 18th International Conference on emerging Networking EXperiments and Technologies}, pages 59--71, 2022.

\bibitem{sharma2023vca}
Taveesh Sharma, Tarun Mangla, Arpit Gupta, Junchen Jiang, and Nick Feamster.
\newblock Estimating webrtc video qoe metrics without using application headers.
\newblock In {\em Proceedings of the 2023 ACM on Internet Measurement Conference}, pages 485--500, 2023.

\bibitem{sirinam2018deep}
Payap Sirinam, Mohsen Imani, Marc Juarez, and Matthew Wright.
\newblock Deep fingerprinting: Undermining website fingerprinting defenses with deep learning.
\newblock In {\em Proceedings of the 2018 ACM SIGSAC conference on computer and communications security}, pages 1928--1943, 2018.

\bibitem{5504793}
Robin Sommer and Vern Paxson.
\newblock Outside the closed world: On using machine learning for network intrusion detection.
\newblock In {\em 2010 IEEE Symposium on Security and Privacy}, pages 305--316, 2010.

\bibitem{tjeng2017evaluatingcertified}
Vincent Tjeng, Kai Xiao, and Russ Tedrake.
\newblock Evaluating robustness of neural networks with mixed integer programming.
\newblock {\em arXiv preprint arXiv:1711.07356}, 2017.

\bibitem{Velan2015ASO}
Petr Velan, Milan Cermák, Pavel Čeleda, and Martin Drašar.
\newblock A survey of methods for encrypted traffic classification and analysis.
\newblock {\em International Journal of Network Management}, 25:355 -- 374, 2015.

\bibitem{wang2023bars}
Kai Wang, Zhiliang Wang, Dongqi Han, Wenqi Chen, Jiahai Yang, Xingang Shi, and Xia Yin.
\newblock Bars: Local robustness certification for deep learning based traffic analysis systems.

\bibitem{Wang2019survey}
Pan Wang, Xuejiao Chen, Feng Ye, and Zhixin Sun.
\newblock A survey of techniques for mobile service encrypted traffic classification using deep learning.
\newblock {\em IEEE Access}, 7:54024--54033, 2019.

\bibitem{wang2021betacertified}
Shiqi Wang, Huan Zhang, Kaidi Xu, Xue Lin, Suman Jana, Cho-Jui Hsieh, and J~Zico Kolter.
\newblock Beta-crown: Efficient bound propagation with per-neuron split constraints for neural network robustness verification.
\newblock {\em Advances in Neural Information Processing Systems}, 34:29909--29921, 2021.

\bibitem{wang2023adversarial}
Yulong Wang, Tong Sun, Shenghong Li, Xin Yuan, Wei Ni, Ekram Hossain, and H~Vincent Poor.
\newblock Adversarial attacks and defenses in machine learning-empowered communication systems and networks: A contemporary survey.
\newblock {\em IEEE Communications Surveys \& Tutorials}, 2023.

\bibitem{wong2020fast}
Eric Wong, Leslie Rice, and J~Zico Kolter.
\newblock Fast is better than free: Revisiting adversarial training.
\newblock {\em arXiv preprint arXiv:2001.03994}, 2020.

\bibitem{xu2020automaticcertified}
Kaidi Xu, Zhouxing Shi, Huan Zhang, Yihan Wang, Kai-Wei Chang, Minlie Huang, Bhavya Kailkhura, Xue Lin, and Cho-Jui Hsieh.
\newblock Automatic perturbation analysis for scalable certified robustness and beyond.
\newblock {\em Advances in Neural Information Processing Systems}, 33:1129--1141, 2020.

\bibitem{Yan2015EnablingQL}
Suying Yan, Yuchun Guo, Yishuai Chen, Feng Xie, Chenguang Yu, and Y.~Liu.
\newblock Enabling qoe learning and prediction of webrtc video communication in wifi networks.
\newblock 2015.

\bibitem{yang2020randomizedcertified}
Greg Yang, Tony Duan, J~Edward Hu, Hadi Salman, Ilya Razenshteyn, and Jerry Li.
\newblock Randomized smoothing of all shapes and sizes.
\newblock In {\em International Conference on Machine Learning}, pages 10693--10705. PMLR, 2020.

\bibitem{yao2021lb}
Zhiyuan Yao, Yoann Desmouceaux, Mark Townsley, and Thomas~Heide Clausen.
\newblock Towards intelligent load balancing in data centers.
\newblock {\em CoRR}, abs/2110.15788, 2021.

\bibitem{netshare-sigcomm2022}
Yucheng Yin, Zinan Lin, Minhao Jin, Giulia Fanti, and Vyas Sekar.
\newblock Practical gan-based synthetic ip header trace generation using netshare.
\newblock In {\em Proceedings of the ACM SIGCOMM 2022 Conference}, SIGCOMM '22, page 458–472, New York, NY, USA, 2022. Association for Computing Machinery.

\end{thebibliography}
